\documentclass[11pt, a4paper]{article}
\usepackage[utf8]{inputenc}
\usepackage{jheppub}
\usepackage{tikz}
\usetikzlibrary{positioning}
\tikzset{font=\small, >=stealth, shorten >=0.1cm, shorten <=0.1cm}

\DeclareFontFamily{U}{MnSymbolC}{}
\DeclareSymbolFont{MnSyC}{U}{MnSymbolC}{m}{n}
\DeclareFontShape{U}{MnSymbolC}{m}{n}{
    <-6>  MnSymbolC5
   <6-7>  MnSymbolC6
   <7-8>  MnSymbolC7
   <8-9>  MnSymbolC8
   <9-10> MnSymbolC9
  <10-12> MnSymbolC10
  <12->   MnSymbolC12}{}
\DeclareMathSymbol{\intprod}{\mathbin}{MnSyC}{'270}

\newcommand{\gf}{\mathfrak{g}}
\newcommand{\hf}{\mathfrak{h}}

\newcommand{\del}{\partial}
\newcommand{\delb}{{\bar\partial}}

\newcommand{\vev}[1]{\langle #1 \rangle}

\renewcommand{\Re}{\mathop{\mathrm{Re}}\nolimits}

\newcommand{\Tr}{\mathop{\mathrm{Tr}}\nolimits}

\newcommand{\SU}{\mathrm{SU}}

\newcommand{\Spin}{\mathrm{Spin}}

\newcommand{\U}{\mathrm{U}}

\newcommand{\iso}{\cong}

\newcommand{\R}{\mathbb{R}}
\newcommand{\C}{\mathbb{C}}

\let\nc\newcommand
\let\renc\renewcommand
\nc{\wbar}{\overline}
\let\td\tilde
\let\wtd\widetilde
\let\wht\widehat
\let\mcl\mathcal

\nc{\ab}{{\bar{a}}} \nc{\at}{\tilde{a}} \nc{\ah}{\hat{a}}
\nc{\bb}{{\bar{b}}} \nc{\bt}{\tilde{b}} \nc{\bh}{\hat{b}}
\nc{\cb}{{\bar{c}}} \nc{\ct}{\tilde{c}} %\nc{\ch}{\hat{c}}
\nc{\db}{{\bar{d}}} \nc{\dt}{\tilde{d}} \renc{\dh}{\hat{d}}
\nc{\eb}{{\bar{e}}} \nc{\et}{\tilde{e}} \nc{\eh}{\hat{e}}
\nc{\fb}{{\bar{f}}} \nc{\ft}{\tilde{f}} \nc{\fh}{\hat{f}}
\nc{\gb}{{\bar{g}}} \nc{\gt}{\tilde{g}} \nc{\gh}{\hat{g}}
\nc{\hb}{{\bar{h}}} \nc{\hh}{\hat{h}} %\nc{\ht}{\tilde{h}}
\nc{\ib}{{\bar{\imath}}} \nc{\ih}{\hat{\imath}} %\nc{\it}{\tilde{\imath}}
\nc{\jb}{{\bar{\jmath}}} \nc{\jt}{\tilde{\jmath}} \nc{\jh}{\hat{\jmath}}
\nc{\kb}{{\bar{k}}} \nc{\kt}{\tilde{k}} \nc{\kh}{\hat{k}}
\nc{\lb}{{\bar{l}}} \nc{\lt}{\tilde{l}} \nc{\lh}{\hat{l}}
\nc{\mb}{{\bar{m}}} \nc{\mt}{\tilde{m}} \nc{\mh}{\hat{m}}
\nc{\nb}{{\bar{n}}} \nc{\nt}{\tilde{n}} \nc{\nh}{\hat{n}}
\nc{\ob}{{\bar{o}}} \nc{\ot}{\tilde{o}} \nc{\oh}{\hat{o}}
\nc{\pb}{{\bar{p}}} \nc{\pt}{\tilde{p}} \nc{\ph}{\hat{p}}
\nc{\qb}{{\bar{q}}} \nc{\qt}{\tilde{q}} \nc{\qh}{\hat{q}}
\nc{\rb}{{\bar{r}}} \nc{\rt}{\tilde{r}} \nc{\rh}{\hat{r}}
\renc{\sb}{{\bar{s}}} \nc{\st}{\tilde{s}} \nc{\sh}{\hat{s}}
\nc{\tb}{{\bar{t}}} \renc{\th}{\hat{t}} %\nc{\tt}{\tilde{t}}
\nc{\ub}{{\bar{u}}} \nc{\ut}{\tilde{u}} \nc{\uh}{\hat{u}}
\nc{\vb}{{\bar{v}}} \nc{\vt}{\tilde{v}} \nc{\vh}{\hat{v}}
\nc{\wb}{{\bar{w}}} \nc{\wt}{\tilde{w}} \nc{\wh}{\hat{w}}
\nc{\xb}{{\bar{x}}} \nc{\xt}{\tilde{x}} \nc{\xh}{\hat{x}}
\nc{\yb}{{\bar{y}}} \nc{\yt}{\tilde{y}} \nc{\yh}{\hat{y}}
\nc{\zb}{{\bar{z}}} \nc{\zt}{\tilde{z}} \nc{\zh}{\hat{z}}

\nc{\Ab}{{\wbar{A}}} \nc{\At}{{\wtd{A}}} \nc{\Ah}{{\wht{A}}}
\nc{\Bb}{{\wbar{B}}} \nc{\Bt}{{\wtd{B}}} \nc{\Bh}{{\wht{B}}}
\nc{\Cb}{{\wbar{C}}} \nc{\Ct}{{\wtd{C}}} \nc{\Ch}{{\wht{C}}}
\nc{\Db}{{\wbar{D}}} \nc{\Dt}{{\wtd{D}}} \nc{\Dh}{{\wht{D}}}
\nc{\Eb}{{\wbar{E}}} \nc{\Et}{{\wtd{E}}} \nc{\Eh}{{\wht{E}}}
\nc{\Fb}{{\wbar{F}}} \nc{\Ft}{{\wtd{F}}} \nc{\Fh}{{\wht{F}}}
\nc{\Gb}{{\wbar{G}}} \nc{\Gt}{{\wtd{G}}} \nc{\Gh}{{\wht{G}}}
\nc{\Hb}{{\wbar{H}}} \nc{\Ht}{{\wtd{H}}} \nc{\Hh}{{\wht{H}}}
\nc{\Ib}{{\bar{I}}} \nc{\It}{{\wtd{I}}} \nc{\Ih}{{\wht{I}}}
\nc{\Jb}{{\bar{J}}} \nc{\Jt}{{\wtd{J}}} \nc{\Jh}{{\wht{J}}}
\nc{\Kb}{{\wbar{K}}} \nc{\Kt}{{\wtd{K}}} \nc{\Kh}{{\wht{K}}}
\nc{\Lb}{{\wbar{L}}} \nc{\Lt}{{\wtd{L}}} \nc{\Lh}{{\wht{L}}}
\nc{\Mb}{{\wbar{M}}} \nc{\Mt}{{\wtd{M}}} \nc{\Mh}{{\wht{M}}}
\nc{\Nb}{{\wbar{N}}} \nc{\Nt}{{\wtd{N}}} \nc{\Nh}{{\wht{N}}}
\nc{\Ob}{{\wbar{O}}} \nc{\Ot}{{\wtd{O}}} \nc{\Oh}{{\wht{O}}}
\nc{\Pb}{{\wbar{P}}} \nc{\Pt}{{\wtd{P}}} \nc{\Ph}{{\wht{P}}}
\nc{\Qb}{{\wbar{Q}}} \nc{\Qt}{{\wtd{Q}}} \nc{\Qh}{{\wht{Q}}}
\nc{\Rb}{{\wbar{R}}} \nc{\Rt}{{\wtd{R}}} \nc{\Rh}{{\wht{R}}}
\nc{\Sb}{{\wbar{S}}} \nc{\St}{{\wtd{S}}} \nc{\Sh}{{\wht{S}}}
\nc{\Tb}{{\wbar{T}}} \nc{\Tt}{{\wtd{T}}} \nc{\Th}{{\wht{T}}}
\nc{\Ub}{{\wbar{U}}} \nc{\Ut}{{\wtd{U}}} \nc{\Uh}{{\wht{U}}}
\nc{\Vb}{{\wbar{V}}} \nc{\Vt}{{\wtd{V}}} \nc{\Vh}{{\wht{V}}}
\nc{\Wb}{{\wbar{W}}} \nc{\Wt}{{\wtd{W}}} \nc{\Wh}{{\wht{W}}}
\nc{\Xb}{{\wbar{X}}} \nc{\Xt}{{\wtd{X}}} \nc{\Xh}{{\wht{X}}}
\nc{\Yb}{{\wbar{Y}}} \nc{\Yt}{{\wtd{Y}}} \nc{\Yh}{{\wht{Y}}}
\nc{\Zb}{{\wbar{Z}}} \nc{\Zt}{{\wtd{Z}}} \nc{\Zh}{{\wht{Z}}}

\nc{\CA}{{\mcl{A}}} \nc{\CAb}{{\wbar{\CA}}} \nc{\CAt}{{\wtd{\CA}}} \nc{\CAh}{{\wht{\CA}}}
\nc{\CB}{{\mcl{B}}} \nc{\CBb}{{\wbar{\CB}}} \nc{\CBt}{{\wtd{\CB}}} \nc{\CBh}{{\wht{\CB}}}
\nc{\CC}{{\mcl{C}}} \nc{\CCb}{{\wbar{\CC}}} \nc{\CCt}{{\wtd{\CC}}} \nc{\CCh}{{\wht{\CC}}}
\nc{\cD}{{\mcl{D}}} \nc{\cDb}{{\wbar{\cD}}} \nc{\cDt}{{\wtd{\cC}}} \nc{\cDh}{{\wht{\cD}}}
\nc{\CE}{{\mcl{E}}} \nc{\CEb}{{\wbar{\CE}}} \nc{\CEt}{{\wtd{\CE}}} \nc{\CEh}{{\wht{\CE}}}
\nc{\CF}{{\mcl{F}}} \nc{\CFb}{{\wbar{\CF}}} \nc{\CFt}{{\wtd{\CF}}} \nc{\CFh}{{\wht{\CF}}}
\nc{\CG}{{\mcl{G}}} \nc{\CGb}{{\wbar{\CG}}} \nc{\CGt}{{\wtd{\CG}}} \nc{\CGh}{{\wht{\CG}}}
\nc{\CH}{{\mcl{H}}} \nc{\CHb}{{\wbar{\CH}}} \nc{\CHt}{{\wtd{\CH}}} \nc{\CHh}{{\wht{\CH}}}
\nc{\CI}{{\mcl{I}}} \nc{\CIb}{{\wbar{\CI}}} \nc{\CIt}{{\wtd{\CI}}} \nc{\CIh}{{\wht{\CI}}}
\nc{\CJ}{{\mcl{J}}} \nc{\CJb}{{\wbar{\CJ}}} \nc{\CJt}{{\wtd{\CJ}}} \nc{\CJh}{{\wht{\CJ}}}
\nc{\CK}{{\mcl{K}}} \nc{\CKb}{{\wbar{\CK}}} \nc{\CKt}{{\wtd{\CK}}} \nc{\CKh}{{\wht{\CK}}}
\nc{\CL}{{\mcl{L}}} \nc{\CLb}{{\wbar{\CL}}} \nc{\CLt}{{\wtd{\CL}}} \nc{\CLh}{{\wht{\CL}}}
\nc{\CM}{{\mcl{M}}} \nc{\CMb}{{\wbar{\CM}}} \nc{\CMt}{{\wtd{\CM}}} \nc{\CMh}{{\wht{\CM}}}
\nc{\CN}{{\mcl{N}}} \nc{\CNb}{{\wbar{\CN}}} \nc{\CNt}{{\wtd{\CN}}} \nc{\CNh}{{\wht{\CN}}}
\nc{\CO}{{\mcl{O}}} \nc{\COb}{{\wbar{\CO}}} \nc{\COt}{{\wtd{\CO}}} \nc{\COh}{{\wht{\CO}}}
\nc{\CP}{{\mcl{P}}} \nc{\CPb}{{\wbar{\CP}}} \nc{\CPt}{{\wtd{\CP}}} \nc{\CPh}{{\wht{\CP}}}
\nc{\CQ}{{\mcl{Q}}} \nc{\CQb}{{\wbar{\CQ}}} \nc{\CQt}{{\wtd{\CQ}}} \nc{\CQh}{{\wht{\CQ}}}
\nc{\CR}{{\mcl{R}}} \nc{\CRb}{{\wbar{\CR}}} \nc{\CRt}{{\wtd{\CR}}} \nc{\CRh}{{\wht{\CR}}}
\nc{\CS}{{\mcl{S}}} \nc{\CSb}{{\wbar{\CS}}} \nc{\CSt}{{\wtd{\CS}}} \nc{\CSh}{{\wht{\CS}}}
\nc{\CT}{{\mcl{T}}} \nc{\CTb}{{\wbar{\CT}}} \nc{\CTt}{{\wtd{\CT}}} \nc{\CTh}{{\wht{\CT}}}
\nc{\CU}{{\mcl{U}}} \nc{\CUb}{{\wbar{\CU}}} \nc{\CUt}{{\wtd{\CU}}} \nc{\CUh}{{\wht{\CU}}}
\nc{\CV}{{\mcl{V}}} \nc{\CVb}{{\wbar{\CV}}} \nc{\CVt}{{\wtd{\CV}}} \nc{\CVh}{{\wht{\CV}}}
\nc{\CW}{{\mcl{W}}} \nc{\CWb}{{\wbar{\CW}}} \nc{\CWt}{{\wtd{\CW}}} \nc{\CWh}{{\wht{\CW}}}
\nc{\CX}{{\mcl{X}}} \nc{\CXb}{{\wbar{\CX}}} \nc{\CXt}{{\wtd{\CX}}} \nc{\CXh}{{\wht{\CX}}}
\nc{\CY}{{\mcl{Y}}} \nc{\CYb}{{\wbar{\CY}}} \nc{\CYt}{{\wtd{\CY}}} \nc{\CYh}{{\wht{\CY}}}
\nc{\CZ}{{\mcl{Z}}} \nc{\CZb}{{\wbar{\CZ}}} \nc{\CZt}{{\wtd{\CZ}}} \nc{\CZh}{{\wht{\CZ}}}

\let\eps\epsilon
\let\ups\upsilon
\let\veps\varepsilon
\let\vtht\vartheta

\let\vsgm\varsigma
\let\vphi\varphi
\let\vrho\varrho

\nc{\alphab}{{\bar{\alpha}}} \nc{\alphat}{{\td{\alpha}}} \nc{\alphah}{{\hat{\alpha}}}
\nc{\betab}{{\bar{\beta}}}   \nc{\betat}{{\td{\beta}}}   \nc{\betah}{{\hat{\beta}}} 
\nc{\gammab}{{\bar{\gamma}}} \nc{\gammat}{{\td{\gamma}}} \nc{\gammah}{{\hat{\gamma}}} 
\nc{\deltab}{{\bar{\delta}}} \nc{\deltat}{{\td{\delta}}} \nc{\deltah}{{\hat{\delta}}} 
\nc{\epsilonb}{{\bar{\eps}}} \nc{\epsilont}{{\td{\eps}}} \nc{\epsilonh}{{\hat{\eps}}} 
\nc{\vepsb}{{\bar{\veps}}}   \nc{\vepst}{{\td{\veps}}}   \nc{\vepsh}{{\hat{\veps}}} 
\nc{\zetab}{{\bar{\zeta}}}   \nc{\zetat}{{\td{\zeta}}}   \nc{\zetah}{{\hat{\zeta}}} 
\nc{\etab}{{\bar{\eta}}}     \nc{\etat}{{\td{\eta}}}     \nc{\etah}{{\hat{\eta}}} 
\nc{\thetab}{{\bar{\theta}}} \nc{\thetat}{{\td{\theta}}} \nc{\thetah}{{\hat{\theta}}} 
\nc{\vthetab}{{\bar{\vtht}}} \nc{\vthetat}{{\td{\vtht}}} \nc{\vthetah}{{\hat{\vtht}}} 
\nc{\lambdab}{{\bar{\lambda}}} \nc{\lambdat}{{\td{\lambda}}} \nc{\lambdah}{{\hat{\lambda}}} 
\nc{\iotab}{{\bar{\iota}}}   \nc{\iotat}{{\td{\iota}}}   \nc{\iotah}{{\hat{\iota}}} 
\nc{\kappab}{{\bar{\kappa}}} \nc{\kappat}{{\td{\kappa}}} \nc{\kappah}{{\hat{\kappa}}} 
\nc{\lmdb}{{\bar{\lmd}}}     \nc{\lmdt}{{\td{\lmd}}}     \nc{\lmdh}{{\hat{\lmd}}} 
\nc{\mub}{{\bar{\mu}}}       \nc{\mut}{{\td{\mu}}}       \nc{\muh}{{\hat{\mu}}} 
\nc{\nub}{{\bar{\nu}}}       \nc{\nut}{{\td{\nu}}}       \nc{\nuh}{{\hat{\nu}}} 
\nc{\xib}{{\bar{\xi}}}       \nc{\xit}{{\td{\xi}}}       \nc{\xih}{{\hat{\xi}}} 
\nc{\pib}{{\bar{\pi}}}       \nc{\pit}{{\td{\pi}}}       \nc{\pih}{{\hat{\pi}}} 
\nc{\vpib}{{\bar{\vpi}}}     \nc{\vpit}{{\td{\vpi}}}     \nc{\vpih}{{\hat{\vpi}}} 
\nc{\rhob}{{\bar{\rho}}}     \nc{\rhot}{{\td{\rho}}}     \nc{\rhoh}{{\hat{\rho}}} 
\nc{\vrhob}{{\bar{\vrho}}}   \nc{\vrhot}{{\td{\vrho}}}   \nc{\vrhoh}{{\hat{\vrho}}} 
\nc{\sigmab}{{\bar{\sigma}}} \nc{\sigmat}{{\td{\sigma}}} \nc{\sigmah}{{\hat{\sigma}}} 
\nc{\vsigmab}{{\bar{\vsgm}}} \nc{\vsigmat}{{\td{\vsgm}}} \nc{\vsigmah}{{\hat{\vsgm}}} 
\nc{\taub}{{\bar{\tau}}}     \nc{\taut}{{\td{\tau}}}     \nc{\tauh}{{\hat{\tau}}} 
\nc{\upsb}{{\bar{\ups}}} \nc{\upst}{{\td{\ups}}} \nc{\upsh}{{\hat{\ups}}} 
\nc{\phib}{{\bar{\phi}}}     \nc{\phit}{{\td{\phi}}}     \nc{\phih}{{\hat{\phi}}} 
\nc{\varphib}{{\bar{\vphi}}}   \nc{\varphit}{{\td{\vphi}}}   \nc{\varphih}{{\hat{\vphi}}} 
\nc{\chib}{{\bar{\chi}}}     \nc{\chit}{{\td{\chi}}}     \nc{\chih}{{\hat{\chi}}} 
\nc{\psib}{{\bar{\psi}}}     \nc{\psit}{{\td{\psi}}}     \nc{\psih}{{\hat{\psi}}} 
\nc{\omegab}{{\bar{\omega}}} \nc{\omegat}{{\td{\omega}}} \nc{\omegah}{{\hat{\omega}}} 

\nc{\Gammab}{{\wbar{\Gamma}}}     \nc{\Gammat}{{\wtd{\Gamma}}}     \nc{\Gammah}{{\wht{\Gamma}}}
\nc{\Deltab}{{\wbar{\Delta}}}     \nc{\Deltat}{{\wtd{\Delta}}}     \nc{\Deltah}{{\wht{\Delta}}}
\nc{\Thetab}{{\wbar{\Theta}}}     \nc{\Thetat}{{\wtd{\Theta}}}     \nc{\Thetah}{{\wht{\Theta}}}
\nc{\Lambdab}{{\wbar{\Lambda}}}   \nc{\Lambdat}{{\wtd{\Lambda}}}   \nc{\Lambdah}{{\wht{\Lambda}}}
\nc{\Xib}{{\wbar{\Xi}}}           \nc{\Xit}{{\wtd{\Xi}}}           \nc{\Xih}{{\wht{\Xi}}}
\nc{\Pib}{{\wbar{\Pi}}}           \nc{\Pit}{{\wtd{\Pi}}}           \nc{\Pih}{{\wht{\Pi}}}
\nc{\Sigmab}{{\wbar{\Sigma}}}     \nc{\Sigmat}{{\wtd{\Sigma}}}     \nc{\Sigmah}{{\wht{\Sigma}}}
\nc{\Upsilonb}{{\wbar{\Upsilon}}} \nc{\Upsilont}{{\wtd{\Upsilon}}} \nc{\Upsilonh}{{\wht{\Upsilon}}}
\nc{\Phib}{{\wbar{\Phi}}}         \nc{\Phit}{{\wtd{\Phi}}}         \nc{\Phih}{{\wht{\Phi}}}
\nc{\Psib}{{\wbar{\Psi}}}         \nc{\Psit}{{\wtd{\Psi}}}         \nc{\Psih}{{\wht{\Psi}}}
\nc{\Omegab}{{\wbar{\Omega}}}     \nc{\Omegat}{{\wtd{\Omega}}}     \nc{\Omegah}{{\wht{\Omega}}}

\newcommand{\rmd}{\mathrm{d}}

\title{\texorpdfstring{$\boldsymbol\Omega$}{Omega}-deformation of B-twisted gauge
  theories and the 3d-3d correspondence}

\author[a]{Yuan Luo,}
\author[a]{Meng-Chwan Tan,}
\author[b,c]{Junya Yagi}
\author[a]{and Qin Zhao}

\emailAdd{yuanluo@nus.edu.sg, mctan@nus.edu.sg, junya.yagi@sissa.it,
  zhaoqin@u.nus.edu}

\affiliation[a]{Department of Physics, National University of
  Singapore \\
2 Science Drive 3, Singapore 117551}

\affiliation[b]{International School for Advanced Studies (SISSA) \\
via Bonomea, 265, 34136 Trieste, Italy}

\affiliation[c]{INFN, Sezione di Trieste \\
via Valerio, 2, 34149 Trieste, Italy}

\abstract{We study $\Omega$-deformation of B-twisted gauge theories in
  two dimensions.  As an application, we construct an
  $\Omega$-deformed, topologically twisted five-dimensional maximally
  supersymmetric Yang--Mills theory on the product of a Riemann
  surface $\Sigma$ and a three-manifold $M$, and show that when
  $\Sigma$ is a disk, this theory is equivalent to analytically
  continued Chern--Simons theory on $M$.  Based on these results, we
  establish a correspondence between three-dimensional $\CN = 2$
  superconformal theories and analytically continued Chern--Simons
  theory.  Furthermore, we argue that there is a mirror symmetry
  between $\Omega$-deformed two-dimensional theories.}


\arxivnumber{1410.1538}

\dedicated{In memory of Tan See Hong.}

\newcommand{\auxF}{\mathbf{F}}
\newcommand{\auxFb}{\overline{\mathbf{F}}}
\newcommand{\auxD}{\mathbf{D}}
\newcommand{\phibt}{\tilde{\phib}}

\begin{document}
\maketitle

\flushbottom

\section{Introduction}

The primary motivation for this work is to better understand the 3d-3d
correspondence via five-dimensional maximally supersymmetric
Yang--Mills (5d MSYM) theory.  In trying to do so, we are naturally
led to study $\Omega$-deformation of B-twisted gauge theories in two
dimensions, which is another theme of the present paper.

The 3d-3d correspondence associates to every three-manifold $M$ an
$\CN = 2$ superconformal field theory $T[M]$ in three dimensions.
(Early works on the subject are \cite{Dimofte:2010tz,
  Terashima:2011qi, Terashima:2011xe, Cecotti:2011iy, Dimofte:2011jd,
  Dimofte:2011ju, Dimofte:2011py}.)  A key fact about $T[M]$ is that
it is closely related to Chern--Simons theory on $M$ with complex
gauge group.  For instance, the partition functions of $T[M]$ on $S^1
\times S^2$ and the squashed three-sphere $S^3_b$ are equal to those
of complex Chern--Simons theory at level $k = 0$ \cite{Yagi:2013fda,
  Lee:2013ida} and $k = 1$ \cite{Cordova:2013cea}, respectively.  More
generally, it has been proposed recently \cite{Dimofte:2014zga} that
the partition function of $T[M]$ on the squashed lens space $L(k,1)_b$
equals that of complex Chern--Simons theory at level $k$.

We are interested in a variant of these relations where $T[M]$ on $S^1
\times_\veps D$ is equated to analytically continued Chern--Simons
theory \cite{Gukov:2003na, Dimofte:2009yn, Witten:2010cx}, which is
the holomorphic part of complex Chern--Simons theory.  Here $S^1
\times_\veps D$ is a twisted product of $S^1$ and a disk $D$, with
parameter $\veps$.  This version is actually more powerful, in the
sense that the partition functions on $S^1 \times_\veps D$ with
various boundary conditions give holomorphic blocks of the theory
\cite{Pasquetti:2011fj, Beem:2012mb}, and the partition function on
$L(k,1)_b$ factorizes into these blocks and their complex conjugates.%
\footnote{This factorization was studied in \cite{Pasquetti:2011fj,
    Beem:2012mb, Taki:2013opa} for $k = 0$, $1$ and proved in
  \cite{Alday:2013lba} for $k = 1$.  The case of general $k$ is
  discussed in \cite{Imamura:2013qxa}.  A similar factorization is
  expected to hold for the partition functions on $L(k,p)_b$
  \cite{Dimofte:2014zga}.}
A derivation of this version was provided by Beem et
al.~\cite{Beem:2012mb}, whose argument built on earlier work of
Witten~\cite{Witten:2010zr, Witten:2011zz}.

One of the main results of this paper is an alternative derivation of
this last relation.  More precisely, we establish the equivalence
between the $Q$-invariant sector of $T[M]$ on $S^1 \times_\veps D$ and
analytically continued Chern--Simons theory on $M$, where $Q$ is a
certain supercharge.

The said equivalence is an example of various correspondences between
theories in $d$ and $6-d$ dimensions that originate from the $\CN =
(2,0)$ superconformal theory in six dimensions.  The best-known among
these is probably the AGT correspondence~\cite{Alday:2009aq,
  Wyllard:2009hg} relating 4d $\CN = 2$ theories and Toda theory,
which one obtains by considering the $(2,0)$~theory compactified and
topologically twisted on Riemann surfaces.  In our case, the
correspondence originates from the $(2,0)$ theory formulated on $S^1
\times_\veps D \times M$, with topological twisting~along the
three-manifold $M$.  The general idea is the following.  When $M$ is
very small, this theory reduces to $T[M]$ on $S^1 \times_\veps D$.  On
the other hand, if one somehow integrates out the degrees of freedom
propagating along $S^1 \times_\veps D$, one should get a theory on
$M$.  The $Q$-invariant sector of the latter is, presumably,
analytically continued Chern--Simons theory.  The correspondence in
question then follows by identifying the two theories coming from the
same 6d theory.

Although the idea may be clear, showing that we indeed get
analytically continued Chern--Simons theory is difficult if we stay
within six dimensions, since the $(2,0)$ theory has no known
Lagrangian description.  To avoid this difficulty, we consider the
limit where the radius $R$ of the $S^1$ is very small.  This allows us
to describe the 6d theory as 5d MSYM theory on $D \times M$, and write
down the Lagrangian explicitly.  Then we can apply localization
techniques to simplify the path integral for correlation functions of
$Q$-invariant operators.  We will show that the path integral for the
5d theory is equivalent to that for analytically continued
Chern--Simons theory, and explain how this result can be used to
establish the claimed equivalence for finite $R$.  The logic of our
argument is essentially the same as those employed in
\cite{Yagi:2013fda, Lee:2013ida} for the $S^1 \times S^2$ case or
\cite{Cordova:2013cea} for the $S^3_b$ case.  (A similar approach was
taken in \cite{Kawano:2012up, Fukuda:2012jr} to establish the
equivalence between a twisted 5d MSYM theory compactified on $S^3$ and
$q$-deformed Yang--Mills theory in two dimensions.)

The construction of the 5d theory is, however, nontrivial and
interesting on its own, and this takes us to the second theme of the
present work.  That is the $\Omega$-deformation of B-twisted gauge
theories.

The nontriviality comes from the fact that we are reducing the 6d
theory on the nontrivial $D$-fibration $S^1 \times_\veps D$ over
$S^1$, constructed by gluing the fiber with a rotation by angle $2\pi
R\veps$.  This rotation induces a deformation of the resulting 5d MSYM
theory on $D \times M$.  To understand what kind of deformation is
induced, suppose we further dimensionally reduce the 5d theory on $M$;
thus, in total, we are reducing the 6d theory on $S^1$ and then on
$M$.  If we interchange the order of reduction, then we would be
reducing a 3d $\CN = 2$ theory on the $S^1$ factor of $S^1
\times_\veps D$.  This would give an $\Omega$-deformed $\CN = (2,2)$
theory on $D$ \cite{Shadchin:2006yz, Dimofte:2010tz}. So going back to
the original order, we find that the 5d MSYM theory we obtain is
deformed in such a way that it becomes an $\Omega$-deformed $\CN =
(2,2)$ gauge theory on $D$ upon dimensional reduction on $M$.  We call
this deformation the $\Omega$-deformation of 5d MSYM theory on $D
\times M$.

For the construction of the $\Omega$-deformed 5d MSYM theory, it is
actually more convenient to generalize $S^1 \times_\veps D$ to $S^1
\times_V \Sigma$, where $\Sigma$ is any Riemann surface, and
$\times_V$ means that the product is twisted with the isometry
$\exp(2\pi RV)$ of $\Sigma$ generated by a Killing vector field $V$.
In this more general setup, we must topologically twist the 6d theory
along $\Sigma$ as well in order to preserve some supersymmetry; then
$Q$ will be a supercharge of the twisted theory that is a scalar on
$\Sigma$ and $M$.  As a result, the $\Omega$-deformed 5d MSYM theory
on $\Sigma \times M$ describing the 6d theory also undergoes
topological twisting along $\Sigma$ (on top of the one along $M$), and
we are interested in the $Q$-invariant sector of this twisted 5d
theory.

In general, $\CN = (2,2)$ gauge theories admit two kinds of
topological twist.  One is the A-twist which uses the vector
R-symmetry $\U(1)_V$, and the other is the B-twist which uses the
axial R-symmetry $\U(1)_A$.  We can see which twist is induced on the
5d theory by considering the case $\Sigma = \R^2$.  It has been
observed that 5d MSYM theory on $\R^2 \times M$ without the
$\Omega$-deformation, viewed as an $\CN = (2,2)$ gauge theory on
$\R^2$, has a superpotential given by the Chern--Simons functional for
a complex gauge field $\CA$ on $M$ \cite{Yagi:2013fda}.  For
nonabelian gauge group, the superpotential is not homogeneous in
$\CA$, and this leads to breaking of $\U(1)_V$.  So the twisting must
be done with $\U(1)_A$.  To summarize, the dimensional reduction of
the $\Omega$-deformed twisted 5d MSYM theory on $M$ is an
$\Omega$-deformed B-twisted gauge theory.  Conversely, we can
construct this 5d theory by ``lifting'' an $\Omega$-deformed B-twisted
gauge theory from two to five dimensions.

Unlike its A-twisted counterpart \cite{Shadchin:2006yz,
  Dimofte:2010tz}, the $\Omega$-deformation of B-twisted gauge
theories has been little studied in the literature.  To achieve our
goal, we should therefore understand it first in a general setup, and
this is what we try to do in section~\ref{Omega-BGT}.  In
\cite{Yagi:2014toa}, the $\Omega$-deformation of B-twisted
Landau--Ginzburg models was formulated, and used to provide a unified
approach to understanding quantization of the integrable system
\cite{Nekrasov:2009rc} and the algebra of supersymmetric loop
operators \cite{Gaiotto:2010be, Ito:2011ea} associated with an $\CN =
2$ gauge theory in four dimensions.  We follow the same strategy as
the one employed there, and formulate the $\Omega$-deformation of
general B-twisted gauge theories.  The construction is relatively
straightforward if the worldsheet $\Sigma$ has no boundary.  In the
situation that $\Sigma$ has a boundary, the supersymmetric action
requires an interesting boundary term which turns out to carry much of
the information on the dynamics of the theory.  We then discuss
boundary conditions, and derive a localization formula for correlation
functions of $Q$-invariant operators, taking $\Sigma = D$.

In section~\ref{Omega-5dMSYM}, we turn to the twisted 5d MSYM theory on
$\Sigma \times M$.  Due to the topological twisting, the theory may be
regarded as a B-twisted gauge theory on $\Sigma$.  Hence, we can
obtain its $\Omega$-deformation by adapting the construction developed
in the previous section.  For $\Sigma = D$, we show that the twisted
theory is equivalent to analytically continued Chern--Simons theory on
$M$ by localization of the path integral, following essentially the
same steps as in the derivation of the 2d localization formula.

We conclude our discussion in section~\ref{3d-3d} by placing the above
results in the context of the 3d-3d correspondence.  We establish the
correspondence between $T[M]$ and analytically continued Chern--Simons
theory described above, and moreover discuss a mirror symmetry between
$\Omega$-deformed $\CN = (2,2)$ theories in two dimensions.

\section{\texorpdfstring{$\boldsymbol\Omega$}{Omega}-deformation of
  B-twisted gauge theories}
\label{Omega-BGT}

In this section we formulate the $\Omega$-deformation of B-twisted
gauge theories in two dimensions, and study general properties of the
deformed theories.  In particular, we derive a localization formula
for correlation functions on a disk.  The construction developed in
this section will be crucial for our discussion in the next section.

\subsection{Supersymmetry transformation laws}

First of all, let us explain what we mean by an $\Omega$-deformation
of a B-twisted theory.  The notion of $\Omega$-deformation was
introduced originally in the context of $\CN = 2$ gauge theories on
$\R^4$ \cite{Moore:1997dj, Lossev:1997bz, Moore:1998et,
  Nekrasov:2002qd, Nekrasov:2003rj}.  The following definition is an
analog in the case of B-twisted gauge theories of a more general
formulation of $\Omega$-deformation \cite{Nekrasov:2010ka}, which
works for topologically twisted $\CN = 2$ gauge theories on arbitrary
four-manifolds admitting isometries.

After the B-twisting, an $\CN = (2,2)$ theory has two supercharges
$\Qb_\pm$ that are scalars on the worldsheet $\Sigma$.  The linear
combination $Q = \Qb_+ + \Qb_-$ satisfies $Q^2 = 0$ up to a central
charge, and is used as the BRST operator of the B-twisted theory.
Given a Killing vector field $V$ on $\Sigma$, an $\Omega$-deformation
with respect to $V$ is a deformation such that the deformed theory has
a BRST operator, which we will still denote by $Q$, satisfying the
deformed relation
\begin{equation}
  \label{Q2=LV}
  Q^2 = L_V.
\end{equation}
Here $L_V$ is the conserved charge acting on fields as the
gauge-covariant Lie derivative $\CL_V$ by $V$.

In order to formulate such a deformation, one can start with a
supergravity theory and try to find a background that realizes the
deformation.  For A-twisted theories on $S^2$, such a supergravity
background was found in \cite{Closset:2014pda}.  In principle, one can
apply a mirror map to this background and obtain the corresponding
deformation for B-twisted theories on $S^2$.  Here we instead follow
the strategy employed in \cite{Yagi:2014toa} for the formulation of
$\Omega$-deformed B-twisted Landau--Ginzburg models.  So let us first
review this strategy.
 
% In \cite{Yagi:2014toa}, an $\Omega$-deformation of B-twisted
% Landau--Ginzburg models was formulated, and this was used to
% elucidate interesting phenomena in $\CN = 2$ gauge theories
% concerning quantization of integrable systems \cite{Nekrasov:2009rc}
% and the algebra of supersymmetric loop operators
% \cite{Gaiotto:2010be, Ito:2011ea}.

 As we have said above, two of the four supercharges of $\CN = (2,2)$
supersymmetry algebra become scalars after the B-twist.  The remaining
two, on the other hand, become components of a one-form supercharge $G
=G_z \rmd z + G_\zb \rmd\zb$.  Suppose $\Sigma = \C$.  Then, these
supercharges are all unbroken, and satisfy the commutation relations
$\{\Qb_-, G_z\} = P_z$ and $\{\Qb_+, G_\zb\} = P_\zb$, where $P = P_z
\rmd z + P_\zb \rmd\zb$ is the generator for translations.  The other
commutators vanish, up to central charges.

Now we pick a Killing vector field $V = V^z \del_z + V^\zb \del_\zb$
and set $Q = \Qb_+ + \Qb_- + \iota_V G$, where $\iota_V$ is the
interior product with $V$.  This operator satisfies $Q^2 = \iota_V P$,
and this is nothing but the $\Omega$-deformed relation \eqref{Q2=LV}
on $\C$.  Hence, $Q$ generates an $\Omega$-deformed supersymmetry
transformation on the flat worldsheet.

What we have to do is to generalize this construction to an arbitrary
choice of $\Sigma$ which is not necessarily flat.  To this end, we
should write down the transformations of fields generated by $Q$ in
the flat case (see e.g.\ \cite{MR2003030} for the standard formulas
for $\CN = (2,2)$ supersymmetry transformations), and rewrite them in
a way that makes sense even when $\Sigma$ is curved.  This is actually
not very hard.

A vector multiplet of the B-twisted supersymmetry consists of a gauge
field $A$, a one-form $\sigma$, and an auxiliary scalar $\auxD$, as
well as fermionic fields which are two scalars $\lambdab_\pm$ and a
one-form $\lambda$.  These are all valued in the Lie algebra $\gf$ of
the gauge group $G$, except that the gauge field is a connection on a
$G$-bundle over $\Sigma$.  To avoid introducing dependence on the
metric on $\Sigma$ to the supersymmetry transformation laws, in our
formulation of the B-twisted gauge theory, we replace $\auxD$ with a
two-form (still called $\auxD$), and $\lambdab_\pm$ by two two-forms
$\alpha$ and $\zeta$; these are related to the original fields by the
Hodge duality, once a metric is chosen.%
\footnote{This replacement is necessary for the metric independence
  even when the $\Omega$-deformation is not present, as can be seen
  from the transformation laws for $\alpha$ and $\zeta$.}
Thus, our vector multiplet consists of a gauge field $A$ and
\begin{equation}
  \sigma \in \Omega^1(\Sigma; \gf), \quad 
  \auxD \in \Omega^2(\Sigma; \gf); \quad
  \lambda \in \Omega^1(\Sigma; \gf), \quad 
  \alpha, \, \zeta \in \Omega^2(\Sigma; \gf).
\end{equation}
By $\Omega^p(\Sigma; \gf)$ we mean the space of $p$-forms in the
adjoint representation.

After some rescaling and shifting of fields, we arrive at the
following $\Omega$-deformed transformation laws for the vector
multiplet:
\begin{equation}
  \label{V-SUSY}
  \begin{aligned}
    \delta A &= i\lambda,
    \\
    \delta\sigma &= \lambda + \iota_V\zeta,
    \\
    \delta\lambda
    &= -i\iota_V F_A + \rmd_A\iota_V\sigma,
    \\
    \delta\zeta
    &= i F_A + \rmd_A\sigma - \sigma \wedge \sigma,
    \\
     \delta\alpha
    &= \rmd_A\sigma + \auxD,
    \\
    \delta \auxD
    &=\rmd_A\iota_V\alpha - [\iota_V\sigma, \alpha]
         - \rmd_A\lambda
         - \lambda \wedge \sigma - \sigma \wedge \lambda
         - \rmd_A\iota_V\zeta.
  \end{aligned}
\end{equation}
Here $\rmd_A = \rmd - iA$ is the gauge-covariant exterior
differential, and $F_A$ is the curvature of~$A$.
% The infinitesimal
%   gauge transformation $\delta_\eps$ generated by $\eps \in \gf$ acts
%   by $\delta_\eps A = \rmd_A\eps$ on the gauge field and $i[\eps,
%   \Psi]$ on an adjoint-valued field $\Psi$.

A chiral multiplet consists of fields valued in a unitary
representation $R$ of $G$, as well as those valued in the complex
conjugate representation $\Rb$ which is isomorphic to the dual
representation.  Those valued in $R$ are a complex scalar $\phi$, a
fermionic one-form $\rho$ and an auxiliary two-form $\auxF$, while
those valued in $\Rb$ are fermionic scalars $\etab$ and $\thetab$.
For the metric independence of supersymmetry transformations, we will
use a two-form $\mub$ instead of $\thetab$.  Thus, the fields in our
chiral multiplet are
\begin{equation}
  \phi \in \Omega^0(\Sigma; R), \quad 
  \auxF \in \Omega^2(\Sigma; R); \quad
  \etab \in \Omega^0(\Sigma; \Rb), \quad 
  \rho \in \Omega^1(\Sigma; R), \quad
  \mub \in \Omega^2(\Sigma; \Rb).
\end{equation}

The $\Omega$-deformed supersymmetry transformation laws for a chiral
multiplet were written down in \cite{Yagi:2014toa} in the case without
coupling to a vector multiplet.  It is straightforward to generalize
the formula to the gauged case:
\begin{equation}
  \begin{aligned}
    \delta\phi &= \iota_V\rho, \\
    \delta\phib &= \etab, \\
    \delta\rho &= \rmd_A\phi - \sigma\phi + \iota_V \auxF, \\
    \delta\etab &= \iota_V\rmd_A\phib + \phib\iota_V \sigma, \\
    \delta\mub &= \auxFb, \\
    \delta\auxF &= \rmd_A\rho - \sigma \wedge \rho + \zeta\phi, \\ 
    \delta\auxFb &= \rmd_A\iota_V\mub + \mub\iota_V\sigma.
  \end{aligned}
\end{equation}

We let $Q$ denote the generator for the supersymmetry transformations.
From the above formulas, one can check that $Q$ squares to $\CL_V =
\iota_V\rmd_A + \rmd_A\iota_V$, modulo the gauge transformation
generated by $i\iota_V\sigma$.%
\footnote{More precisely, the supersymmetry transformation laws only
  show that $Q$ obeys $Q^2 = L_V$ if its action is restricted to
  fields.  Actually, on the right-hand side of this relation, an extra
  operator may be present that commutes with any fields.  Such an
  operator corresponds to a central charge in the $\CN=(2,2)$
  supersymmetry algebra.  We will not consider this possibility since
  our discussion only concerns the action of $Q$ on fields.}
Observables are gauge- and $Q$-invariant operators that are not
$Q$-exact.  From the supersymmetry transformation laws, we see that
gauge-invariant functions of $\phi$, inserted at zeros of $V$, are
local observables.

\subsection{Supersymmetric action}
\label{SUSY-action}

Let us construct an action that is invariant under the
$\Omega$-deformed supersymmetry transformations.  It takes the form
\begin{equation}
  S = S_V + S_C + S_W.
\end{equation}
The first two pieces $S_V$ and $S_C$ contain kinetic terms for the
vector and chiral multiplets, respectively, and the last piece $S_W$
contains terms constructed from a superpotential $W$, a
gauge-invariant holomorphic function of the chiral multiplet scalar
$\phi$.

To construct $S_V$ and $S_C$, we need to pick a complex structure and
a K\"ahler metric on $\Sigma$.  We denote the K\"ahler metric by $h$.
Then, the vector multiplet action is
\begin{equation}
  \label{S_V}
  \begin{split}
    S_V
    & = \delta\int_\Sigma \Tr\bigl(
           \alpha \wedge \star(-\rmd_A\sigma + \auxD + 4\delb_A\sigma)
           + \zeta \wedge \star (-iF_A + \rmd_A\sigma + \sigma \wedge \sigma)\bigr)
           \\
    & =\int_\Sigma \Tr\Bigl(
          F_A \wedge \star F_A
          + \sigma \wedge \star \Delta \sigma
          + \frac{\kappa}{2} \sigma \wedge \star\sigma
          - (\sigma \wedge \sigma) \wedge \star(\sigma \wedge \sigma)
          \\ & \qquad 
          + \auxD' \wedge \star\auxD'
          + 2\del_A(\sigma \star\delb_A\sigma) 
          + 2\delb_A(\sigma \star\del_A\sigma) 
          \\  & \qquad 
          - 2\alpha \wedge \star
               \rmd_{A - i\sigma} (\lambda^{1,0} - \lambda^{0,1})
          - 2\zeta \wedge \star\rmd_{A + i\sigma} \lambda
          \\ & \qquad 
          - \alpha \wedge \star\rmd_{A - i\sigma} \iota_V\alpha
          - \zeta \wedge \star\rmd_{A + i\sigma} \iota_V \zeta
          - 2\alpha \wedge \star
               \rmd_A\bigl((\iota_V\zeta)^{1,0} - (\iota_V\zeta)^{0,1}\bigr)
          \Bigr),
  \end{split}
\end{equation}
and the chiral multiplet action is
\begin{equation}
  \label{S_C}
  \begin{split}
    S_C
    & = \delta\int_\Sigma \bigl(
           \rho \wedge \star (\rmd_A\phib - \phib\sigma + \iota_V \auxFb)
           - i\phi\phib\alpha
           + \auxF\wedge\star\mub
           + 2\sigma\phi \wedge \star\iota_V\mub\bigr)
   \\
   & = \int_\Sigma \bigl(
          (\rmd_A\phi + \iota_V \auxF)
          \wedge \star (\rmd_A\phib + \iota_V \auxFb)
          + \sigma\phi \wedge \star(\phib\sigma)
          - i\phi\phib\auxD' 
          + \auxF \wedge \star \auxFb \\
          & \qquad
          - \iota_V \auxF \wedge \star(\phib\sigma)
          + \sigma\phi \wedge \star\iota_V\auxFb
          - i\del(\phi\phib\sigma) + i\delb(\phi\phib\sigma) \\
          & \qquad
          - \rho \wedge \star\rmd_{A - i\sigma} \etab
          + \rmd_{A - i\sigma} \rho \star\mub
          + 2\rho\phib \wedge \star\lambda
          - i\phi\etab\alpha
          + \zeta\phi\star\mub
          \\ &\qquad 
          + \rho \wedge \star(\phib\iota_V\zeta
               - \iota_V\rmd_A\iota_V\mub
               - \iota_V\mub \iota_V\sigma)
          + 2(\lambda\phi + \iota_V\zeta\phi + \sigma\iota_V\rho)
             \wedge \star\iota_V\mub\bigr).
          % + \rho\phib \wedge \star\iota_V\zeta
          % - i\iota_V\rho\phib\alpha
          % - \rho\wedge \iota_V\rmd_{A + \sigma} \iota_V\mub
          % + 2\iota_V\zeta\phi \wedge \star\iota_V\mub\bigr).
  \end{split}
\end{equation}
Here $\Delta = D^* D$ is the Laplacian associated to the covariant
derivative $D$ coupled to the gauge field and the Levi-Civita
connection, $\kappa$ is the scalar curvature, and $\auxD' = \auxD +
2\delb_A\sigma$ is a redefined auxiliary field.

Both $S_V$ and $S_C$ are $Q$-invariant, provided that $V$ is a Killing
vector field.  This follows from the fact that $\CL_V$ commutes with
the Hodge star operator $\star$ for such $V$; thanks to this property,
we have
\begin{equation}
  \delta^2 \int_\Sigma \CV
  = \int_\Sigma \CL_V \CV
  = \int_{\del \Sigma} \iota_V\CV
\end{equation}
for any gauge-invariant two-form $\CV$ on $\Sigma$ constructed from
fields using $\star$, and the last expression vanishes since
$\iota_V\CV$ restricts to zero on $\del\Sigma$, with $V$ being tangent
to $\del\Sigma$.

The construction of the superpotential term is a little tricky if
$\Sigma$ has a boundary.  For simplicity, we will assume that $\Sigma$
has only a single connected boundary component.%
\footnote{If $\Sigma$ has multiple boundary components, then for each
  component one has a boundary term similar to the one in the formula
  \eqref{S_W}.}
The boundary is topologically a circle, and we can choose a periodic
coordinate $\vphi$ (with period $2\pi$) on the boundary such that
\begin{equation}
  \label{veps}
  V|_{\del\Sigma} = \veps \del_{\vphi}
\end{equation}
for some real $\veps$.  Furthermore, we assume that $V$ generates
nontrivial isometries on the boundary, that is, $\veps \neq 0$.  Then
\begin{equation}
  \label{S_W}
  S_W
  = i\int_\Sigma \Bigl(
     \auxF \frac{\del W}{\del\phi}
     + \frac{1}{2} \rho \wedge \rho \frac{\del^2 W}{\del\phi\del\phi}
     + \auxFb \frac{\del \Wb}{\del\phib}
     + \etab \mub \frac{\del^2\Wb}{\del\phib\del\phib}\Bigr)
     - \frac{i}{\veps}
        \int_{\del\Sigma} W \, \rmd \vphi,
\end{equation}
where contraction of gauge indices is implicit.  The boundary term is
needed for $Q$-invariance.

We impose the reality condition such that $\sigma$ and $\auxD'$ are
hermitian, while $\phi^\dagger = \phib$ and $\auxF^\dagger = \auxFb$,
so that the real part of the action is nonnegative in the absence of
boundary.%
\footnote{This is true even when $\kappa < 0$ since the bosonic part
  of $S_V$ can be written as the integral of $\Tr((F_A + i\sigma
  \wedge \sigma) \wedge \star (F_A + i\sigma \wedge \sigma) +
  4\del_A\sigma \wedge \star\delb_A\sigma + \auxD' \wedge
  \star\auxD')$, which is manifestly nonnegative.}
If $\Sigma$ has a boundary, we should impose a suitable boundary
condition on $\phi$ in order to ensure the convergence of the path
integral.

One of the most important features of the action constructed above is
that although it depends on the complex structure and the metric of
$\Sigma$, the dependence is $Q$-exact.  Still, the $\Omega$-deformed
B-twisted theory is not quite topological.  Rather, it is
quasi-topological, in the sense that it is invariant under
deformations of the complex structure and the metric as long as $V$
remains as a Killing vector field.

So far $V$ has been assumed to be a real vector field.  We can relax
this condition and multiply $V$ by a phase factor, since the action
remains $Q$-invariant and nonnegative if we simply replace the
appearance of $V$ by its complex conjugate $\Vb$ in the first line of
the formula \eqref{S_C} for $S_C$.  A phase rotation of $V$ is
actually equivalent to the opposite phase rotation of $W$, for the
former has the same effect as the latter combined with the action of
an element in the vector R-symmetry group $\U(1)_V$ (with the chiral
multiplet assigned charge $0$ under it), but the $\U(1)_V$-action can
be undone by a field redefinition (which does not modify the path
integral measure, as there is no quantum anomaly for $\U(1)_V$).

\subsection{Boundary condition}
\label{BC}

We have constructed the $\Omega$-deformed B-twisted theory on a
general worldsheet $\Sigma$.  In particular, we allowed the
possibility that $\Sigma$ has a boundary.  We now discuss boundary
conditions.

The boundary of $\Sigma$ is topologically a circle, and the Killing
vector field $V$ generates its rotations.  The neighborhood of the
boundary looks like a short cylinder.  We equip this cylinder with a
flat metric $\rmd s^2 = \rmd n^2 + \rmd\vphi^2$, with $n$ being a
coordinate in the direction normal to the boundary.  After the
boundary condition is fixed, one can deform the metric of $\Sigma$ to
anything that is allowed by the quasi-topological property of the
theory.  However, the boundary condition will depend on the initial
choice of the flat metric in the neighborhood of the boundary.

Our boundary conditions must meet two requirements.  One is that they
should lead to a good variational problem in a semiclassical, or weak
coupling, limit.  In our case there is a natural weak coupling limit,
which is obtained by rescaling the $Q$-exact part of the action by a
large factor; correlation functions of $Q$-invariant operators are
left unchanged under such a $Q$-exact deformation.  So we require that
boundary terms be absent in the variation of the action when we vary
the fields in this limit.  The other requirement is that boundary
conditions must be $Q$-invariant so that $Q$ preserves the space of
allowed field configurations.

We first analyze boundary conditions for the vector multiplet fields.
The gauge field has the standard kinetic term, so its boundary
condition is a standard one, namely either the Dirichlet or Neumann
boundary condition.  Since a gauge-invariant expression for the former
condition does not exist in two dimensions, we choose the latter,
$F_{n\vphi} = 0$.  Gauging $A_n$ away, we can write this condition as
$\del_n A_\vphi = 0$.  The requirement of $Q$-invariance then leads to
$\del_n\sigma_\vphi = \lambda_n = \del_n \lambda_\vphi = 0$.  If we
now look at the kinetic term for $\sigma$ in the vector multiplet
action \eqref{S_V}, we notice that it differs from the standard one by
total derivative terms.  A natural way to kill these unwanted terms is
to set $\sigma_n = 0$ on the boundary; the total derivative terms in
the chiral multiplet action~\eqref{S_C} also drop out then.  Taking
the $Q$-variation of this condition, we get $\zeta_{n\vphi} = 0$.

In fact, the set of boundary conditions we have found so far is part
of the conditions imposed by a B-brane in $\CN = (2,2)$ gauge theory
\cite{Hori:2013ika, Herbst:2008jq}.  This suggests that we should
choose our boundary condition for the vector multiplet to be the
B-brane condition:
\begin{equation}
  \label{BC-V}
  A_n = \del_n A_\vphi = \sigma_n = \del_n\sigma_\vphi
  = \lambda_n = \del_n \lambda_\vphi = \zeta_{n\vphi}
  = \del_n\alpha_{n \vphi} = \del_n\auxD'_{n\vphi} = 0.
\end{equation}
This set of boundary conditions is not $Q$-invariant by itself.  In
order to achieve $Q$-invariance, we further impose an infinite series
of conditions, generated from the above conditions by the action of
even powers of $\del_n$ \cite{Hori:2013ika}.

We stress that the gauge $A_n = 0$ has been chosen on the boundary
above.  For compatibility, we must restrict gauge transformations to
be such that their parameters have vanishing normal derivatives on the
boundary.  In addition, we can impose a restriction on the boundary
values of gauge transformations.  To do so, we pick a subgroup $H$ of
$G$ and require gauge transformations to be valued in $H$ on the
boundary.  Then, those gauge transformations that do not satisfy this
condition form a physical symmetry of the theory, provided that they
leave invariant the boundary condition for the chiral multiplet, to
which we now turn.

Thanks to the condition $A_n = 0$, the boundary terms arising from
variation of the chiral multiplet action~\eqref{S_C} are all
independent of the vector multiplet fields.  Furthermore, the
supersymmetry variations for $\phi$ and $\phib$ do not depend on
vector multiplet fields either.  In this situation, the analysis of
the boundary condition for the chiral multiplet reduces to the case of
Landau--Ginzburg models \cite{Yagi:2014toa}.  Hence, we can impose the
same boundary condition as in that case.

We refer the reader to \cite{Yagi:2014toa} for the details of the
analysis, and here simply state the result.  The boundary condition
for the chiral multiplet depends on a choice of a submanifold $\gamma$
in the target space, which may be considered as the support of a brane
of a certain type.  Then the scalars obey the usual D-brane boundary
condition:
\begin{equation}
  \label{BBC}
  \phi \in \gamma, \quad
  \del_n\phi \in N_\R\gamma
\end{equation}
at each point on $\del\Sigma$, where $N_\R\gamma$ is the normal bundle
of $\gamma$.  The fermions obey
\begin{equation}
  \label{FBC}
  (\veps\rho_\vphi, \etab) \in T_\C\gamma, \quad
  (\veps\rho_n, \mub_{n\vphi}) \in N_\C\gamma.
\end{equation}
Again, there are further conditions obtained by repeated action of $Q$
on the above conditions, which guarantee that the boundary condition
is $Q$-invariant.

The target space for the chiral multiplet is the representation space
$V_R$ of $R$, equipped with the $G$-invariant K\"ahler form
\begin{equation}
  \omega = i\rmd\phi \wedge \rmd\phib.  
\end{equation}
Note that here $\phi = (\phi^1, \dotsc, \phi^{\dim V_R})$ is
considered as a set of complex coordinates on $V_R$; thus $\rmd\phi$
is a set of $(1,0)$-forms on $V_R$.  We require $\gamma$ to be a
Lagrangian submanifold of the symplectic manifold $(V_R, \omega)$.  As
we will see, this has the effect of eliminating fermion zero modes.

Moreover, $\gamma$ must be $H$-invariant for gauge invariance to be
unbroken.  This requirement has the following consequence.  Let
$\{T_a\}$ be a set of generators of $G$, and $X_a$ denote the vector
fields on $V_R$ generated by the action of $T_a$.  The moment map
$\mu\colon V_R \to \gf^\vee$ for the $G$-action on the symplectic
manifold $(V_R, \omega)$ is given by
\begin{equation}
  \label{mu}
  (\mu, T_a) = i\phi\phib T_a.
\end{equation}
Let $\mu_H\colon V_R \to \hf^\vee$ be the moment map for the
$H$-action; by definition, $(\rmd\mu_H, T_a) = \iota_{X_a}\omega$ for
$T_a \in \hf$.  Since $\gamma$ is $H$-invariant and a Lagrangian
submanifold by assumption, $X_a$ are tangent to $\gamma$ and
$\iota_{X_a}\omega$ vanishes on $T\gamma$.  It follows that $\mu_H$ is
constant on the boundary, as any variation of $\phi$ is tangent to
$\gamma$ due to the boundary condition $\phi \in \gamma$.

\subsection{Localization}
\label{localization}

Finally, we derive a formula for correlation functions of
$Q$-invariant operators via localization of the path integral.  We
take our worldsheet $\Sigma$ to be a disk $D$, and equip it with a
rotationally invariant metric $h$.  The Killing vector field $V$
generates rotations and can be written as $V = \veps\del_\vphi$ for
some $\veps \neq 0$.  By the quasi-topological property of the theory,
we can always deform $h$ into a metric with scalar curvature $\kappa >
0$, such as one for a hemisphere.  We choose the subgroup $H$ to be
trivial, that is, we choose to divide the field space by gauge
transformations that equal the identity on the boundary.

In order to localize the path integral, one usually rescales the
$Q$-exact part of the action by a large factor $t$, which in our case
means rescaling $S_V + S_C \to t(S_V + S_C)$.  On the other hand, we
expect that the theory simplifies considerably when $D$ is very small,
since in such a situation most degrees of freedom are very massive and
decouple from the dynamics.  So we may also want to rescale the metric
as $h \to t^{-2} h$.  If we combine these two ways to simplify the
path integral, the net effect is that $S_V$ is rescaled by a factor of
$t^3$, while $S_C$ is rescaled by a factor of $t$, except the term
coming from $F \wedge \star\mub$ in the $Q$-exact expression
\eqref{S_C} which is rescaled by $t^3$.  Motivated by this
consideration, we deform the action as follows:
\begin{equation}
  S \to
  t^3 \Bigl(S_V
  - \delta\int_D \Tr\alpha \star\auxD'\Bigr)
   + t\Bigl(S_C
  + s \delta\int_D \auxF \star\mub\Bigr)
  + S_W.
\end{equation}
Here $s$ is a real parameter.

First, we rescale $\mub \to s^{-1}\mub$ and take the limit $s \to
\infty$.  In this limit, integrating out the auxiliary field $\auxF$
is equivalent to simply setting
\begin{equation}
  \auxF = 0.  
\end{equation}
The term containing $\auxD'$ is included in the deformation so that
integrating $\auxD'$ out produces delta functions imposing the
constraint
\begin{equation}
  \label{mu=Dsigma}
  \mu = -t^2 \star(\del_A\sigma - \delb_A\sigma),
\end{equation}
where the moment map $\mu$ is given by the formula \eqref{mu}, and
$\sigma$ is regarded as valued in $\gf^\vee$ by $(\sigma, X) =
\Tr(\sigma X)$ for $X \in \gf$.

Next, we take $t$ to be large (but still finite).  Looking at the
bosonic parts of $S_V$ and $S_C$, we find that the path integral then
localizes, under the boundary condition \eqref{BC-V}, to the locus
given by
\begin{equation}
  F_A = \sigma = \rmd_A\phi = 0.
\end{equation}
As $D$ is simply connected, the equation $F_A = 0$ means that we can
set
\begin{equation}
  A = 0
\end{equation}
everywhere by a gauge transformation.  Together with the boundary
condition \eqref{BBC} and the constraint \eqref{mu=Dsigma}, the
equations $\sigma = \rmd_A\phi = 0$ then imply that the path integral
localizes to the configurations where $\phi$ is a constant map to the
subspace $\gamma \cap \mu^{-1}(0)$ of the target space $V_R$.
% These configurations satisfy the boundary condition described in
% section~\ref{BC}.

Since the path integral localizes for large $t$, we can evaluate it by
perturbation theory (in $1/\sqrt{t}$) around background configurations
on the localization locus.  We will denote backgrounds with subscript
$0$ and fluctuations around them with a tilde; thus $A = \At$, \
$\sigma = \sigmat$, and $\phi = \phi_0 + \phit$.

For the computation we need to fix the gauge.  We choose the standard
gauge-fixing condition $\nabla_\mu A^\mu = 0$ and add to the action the
gauge-fixing term
\begin{equation}
  S_G
  = t^3 \int_{D} \sqrt{h} \, \rmd^2 x \Tr\bigl(
     \cb \nabla^\mu D_\mu c+ (\nabla^\mu A_\mu)^2\bigr),
\end{equation}
where $\nabla$ is the Levi-Civita connection and $c$, $\cb$ are
ghosts.  After rescaling the fluctuations and the fermions as
\begin{equation}
  \label{rescaling}
  \begin{aligned}
    (\At, \sigmat, \lambda, \zeta, \alpha)
    &\to (t^{-3/2}\At,  t^{-3/2}\sigmat,
             t^{-1}\lambda, t^{-2}\zeta, t^{-2}\alpha),
    \\
    (\phit, \rho, \etab, \mub)
    &\to (t^{-1/2} \phit, t^{-1}\rho, \etab, \mub),
    \\
    (c, \cb)
    &\to (t^{-3/2} c, t^{-3/2} \cb),
  \end{aligned}
\end{equation}
the terms in the action containing them become
\begin{multline}
  \label{quad-terms}
  \int_{D}  \bigl(\Tr(\At \wedge \star\Delta_\rmd\At
               + \sigmat \wedge \star\Delta_\rmd\sigmat
               + \alpha \wedge \star(\del\lambda - \delb\lambda) 
               - 2\zeta \wedge \star\rmd\lambda 
               + \cb \wedge \star\Delta_\rmd c) \\
  + \rmd\phit \wedge \star\rmd\phibt
  - \rho \wedge\star\rmd\etab
  + \rmd\rho \wedge \star\mub  + \dotsb\bigr).
\end{multline}
Here $\Delta_\rmd = (\rmd + \rmd^*)^2$ is the Hodge--de Rham
laplacian, and the ellipsis refers to terms multiplied by negative
powers of $t$.  To obtain this expression we have used the relation
$\Delta_\rmd = \nabla^* \nabla + \kappa/2$ in the space of one-forms
on a surface.

We have to integrate over the fluctuations and the fermions.  To do
this, we deform $D$ into the shape of a two-sphere $S^2$ with a small
disk $D_\eps$ of radius $\eps$ removed.  Since fields on $S^2
\setminus D_\eps$ can be obtained from fields on $S^2$ by restriction,
we can expand them in the eigenmodes of $\Delta_\rmd$ on $S^2$.%
\footnote{On $S^2$, the fermionic part of the leading terms in the
  integral \eqref{quad-terms} can be written as
  \begin{equation}
    -\langle\rho, (\rmd + \rmd^*)(\etab + \mub)\rangle
    + \langle\alpha - 2\zeta, (\rmd + \rmd^*) \lambda^{0,1}\rangle
    - \langle\alpha + 2\zeta, (\rmd + \rmd^*) \lambda^{1,0}\rangle
    + \langle\cb, \Delta_\rmd c\rangle,
  \end{equation}
  using an appropriate inner product $\langle\, , \,\rangle$.  It is
  thus natural to expand the fermions in the eigenmodes of
  $\Delta_\rmd$.}
The integral~\eqref{quad-terms}, when expressed in terms of the
expansion coefficients, differs from the case with $\Sigma = S^2$ by
$\eps$-dependent terms.  However, at the end of the localization
computation, we can take the limit $\eps \to 0$ (which is a $Q$-exact
operation), whereby the difference simply vanishes.  Thus, we can
perform the integration over the fluctuations and the fermions in a
way similar to the $S^2$ case.  The computation is not quite like that
case, however, since the boundary condition imposes relations among
the expansion coefficients.

To understand the result of the integration, we note the following
three points.  First, the leading terms in the integral
\eqref{quad-terms} are completely independent of the background.
Second, the boundary condition does not depend on the background
either.  This is because the support $\gamma$ of the brane is a
Lagrangian submanifold of $V_R$, and the tangent bundles at different
points on $\gamma$ are all isomorphic up to unitary rotations which
are symmetries of the action.  Finally, there are no fermion zero
modes, as we will see shortly; they are all eliminated by the boundary
condition.  Hence, to leading order, the integration over the
fluctuations and the fermions just produces a constant independent of
the background, though it may depend on the choice of $\gamma$ and the
representation $R$.

Once the perturbative computation is carried out, we integrate over
the localization locus.  On this locus, the only surviving piece of
the action is the boundary term in the superpotential term
\eqref{S_W}:
\begin{equation}
  - \frac{i}{\veps} \int_{\del D} W \, \rmd \vphi
  = -\frac{2\pi i}{\veps} W.
\end{equation}
Finally, taking the limit $t \to \infty$ whereby the subleading terms
vanish, we conclude that the correlation function on the disk of any
$Q$-invariant operator $\CO$ is given by the formula
\begin{equation}
  \label{loc-formula}
  \vev{\CO}
  = \int_{\gamma \cap \mu^{-1}(0)} \rmd\phi_0
     \exp\Bigl(\frac{2\pi i}{\veps} W\Bigr) \CO.
\end{equation}
From this formula we see that the nontrivial information on the
dynamics on $D$ is encoded in the boundary term.

In the above derivation, we have asserted that there are no fermion
zero modes.  Let us show this now.  Recall that we have expanded the
fermions in the eigenmodes of the Laplacian on $S^2$.  There are no
harmonic one-forms on $S^2$, so there are no zero modes for $\lambda$
and $\rho$.  Furthermore, harmonic two-forms are Hodge duals of
constants, and neither constant $\star\alpha$ nor $\star\zeta$ is
compatible with the boundary condition $\del_n \alpha_{n\vphi} =
\zeta_{n\vphi} = 0$.  (To see this for $\alpha$, suppose that we equip
$D$ with the metric $(\rmd n^2 + n^2 \rmd\vphi^2)/(1 + n^2)^2$ of the
Riemann sphere parametrized by $z = ne^{i\vphi}$, where $(n,\vphi)$
are the cylindrical coordinates used in describing the boundary
condition, with the boundary located at $n = 0$.  Then, the zero mode
of $\alpha$ behaves as $\alpha_{n\vphi} \sim n/(1 + n^2)^2$ near the
boundary.)  The boundary condition \eqref{FBC}, on the other hand,
implies that the zero mode parts $\etab_0$, $\mub_0$ of $\etab$,
$\mub$, obey $(0, \etab_0) \in T_\C\gamma$ and $(0, \star\mub_0) \in
N_\C\gamma$ on the boundary.%
\footnote{To be precise, the boundary condition is imposed on the
  fermionic fields themselves and not just on their zero modes.
  However, if we take the limit such that the radius of the $S^2$ goes
  to zero, all nonzero modes become infinitely massive and decouple.
  Then the fermions may be replaced by their zero mode parts, and the
  boundary condition is written entirely in terms of the zero modes.}
Since $\gamma$ is a Lagrangian submanifold of a K\"ahler manifold
for which the complex structure exchanges the tangent and normal
bundles, it follows that $\etab_0 = \mub_0 = 0$ on the boundary and
hence everywhere.  So there are no zero modes for $\etab$ and $\mub$,
either.  Lastly, the zero modes for the ghosts $c$, $\cb$ are
constant, but there are no such modes to begin with.  This is a
consequence of our choice to divide the field space by gauge
transformations that equal the identity on the boundary; gauge
transformation parameters must vanish on the boundary, therefore so do
the ghosts.

\section{Chern--Simons theory from 5d MSYM theory}
\label{Omega-5dMSYM}

As an application of the formulation developed above, in this section
we construct an $\Omega$-deformation of 5d MSYM theory, placed and
topologically twisted on $\Sigma \times M$, where $M$ is a
three-manifold.  This is achieved by ``lifting'' the supersymmetry
transformation laws and the supersymmetric action constructed in the
previous section from two to five dimensions.  Then, we show that when
$\Sigma$ is a disk $D$, the $\Omega$-deformed twisted 5d MSYM theory
is equivalent to analytically continued Chern--Simons theory on $M$,
with integration contour specified by the boundary condition of the 5d
theory.  These results will be the bases for our derivations of
various correspondences presented in the next section.

\subsection[\texorpdfstring{$\Omega$}{Omega}-deformed twisted 5d MSYM
theory on \texorpdfstring{$\Sigma \times M$}{Sigma x
  M}]{$\boldsymbol\Omega$-deformed twisted 5d MSYM theory on
$\boldsymbol{\Sigma \times M}$}

To begin, we formulate the $\Omega$-deformation of the twisted 5d MSYM
theory on $\Sigma \times M$.  The gauge group is a compact Lie group
$G$.  We equip the Riemann surface $\Sigma$ with a metric $h_\Sigma$
and $M$ with a metric $h_M$, and choose a Killing vector field $V$
generating isometries of $\Sigma$.  The metric on $\Sigma \times M$ is
$h = h_\Sigma \oplus h_M$.  We write $(x^M) = (x^\mu, x^m)$ for
coordinates on $\Sigma \times M$.

The theory is topologically twisted as follows.  The structure group
of the spinor bundle of $\Sigma \times M$ is $\Spin(2)_\Sigma \times
\Spin(3)_M \iso \U(1)_\Sigma \times \SU(2)_M$.  Correspondingly, we
split the R-symmetry group $\Spin(5)_R$ as $\Spin(2)_R \times
\Spin(3)_R \iso \U(1) _R \times \SU(2)_R$.  The field content of the
untwisted theory consists of a gauge field $A$, five scalars $X$, and
fermions $\Psi$, transforming under $\SU(2)_M \times \SU(2)_R \times
\U(1)_\Sigma \times \U(1)_R$ as
\begin{equation}
  \begin{aligned}
    A &\colon  (\mathbf{1}, \mathbf{1})^{(\pm 2,0)}
                     \oplus (\mathbf{3}, \mathbf{1})^{(0,0)},
    \\
    X &\colon (\mathbf{1}, \mathbf{1})^{(0,\pm  2)}
                     \oplus (\mathbf{1}, \mathbf{3})^{(0,0)},
    \\
    \Psi &\colon (\mathbf{2}, \mathbf{2})^{(\pm 1,\pm 1)}.
  \end{aligned}
\end{equation}
First, we replace $\SU(2)_M$ with the diagonal subgroup
$\SU(2)_M'$ of $\SU(2)_M \times \SU(2)_R$.  Under $\SU(2)_M' \times
\U(1)_\Sigma \times \U(1)_R$, the fields transform as
\begin{equation}
  \begin{aligned}
    A &\colon \mathbf{1}^{(\pm 2, 0)} \oplus \mathbf{3}^{(0,0)},
    \\
    X &\colon \mathbf{1}^{(0, \pm 2)} \oplus \mathbf{3}^{(0,0)},
    \\
    \Psi &\colon \mathbf{1}^{(\pm 1, \pm 1)}
                         \oplus \mathbf{3}^{(\pm 1, \pm 1)}.
  \end{aligned}
\end{equation}
From the transformation property of $\Psi$, we see that the theory now
has $\CN = (2,2)$ supersymmetry on $\Sigma$.  Next, we identify
$\U(1)_R$ with the axial R-symmetry group $\U(1)_A$, and perform the
B-twist on $\Sigma$, replacing $\U(1)_\Sigma$ with the diagonal
subgroup of $\U(1)_\Sigma \times \U(1)_R$.

In the language of $\CN = (2,2)$ supersymmetry on $\Sigma$, the fields
of the twisted 5d MSYM theory are grouped into a vector multiplet that
is a scalar on $M$, and three adjoint-valued chiral multiplets that
combine into a one-form on $M$.  (Recall, however, that some of the
fermions are redefined in our construction.)  The scalars of the
chiral multiplets are complex combinations of the components $A_m$ of
$A$ along $M$ and three scalars $X_m$:
\begin{equation}
  \CA_m = A_m + iX_m, \quad
  \CAb_m = A_m - iX_m 
\end{equation}
These can be regarded as components of a complex gauge field $\CA =
\CA_m \rmd x^m$ on $M$ and its hermitian conjugate $\CAb$.

Being a B-twisted gauge theory, the $\Omega$-deformation of the
twisted 5d MSYM theory can be formulated in a way similar to the
construction discussed in the previous section.  To adapt, or
``lift,'' that construction to the present 5d setup, we just need to
replace every appearance of $-iA_m$ in our formulas with the covariant
derivative $D_m = \nabla_m - iA_m$ with respect to $A_m$ and the
Levi-Civita connection on $M$; the replacement makes the formulas
invariant under 5d gauge transformations, and provides derivatives
along $M$.  Actually, only the combinations $-i\CA_m$ and $-i\CAb_m$
appear, and these are replaced with $\cD_m = D_m + X_m$ and $\cDb_m =
D_m - X_m$, respectively.
% This lifting construction was discussed in \cite{Yagi:2013fda} in
% the context of 5d MSYM theory on $S^2 \times M$.

For those fields that are in the vector multiplet on $\Sigma$, the
lifted supersymmetry transformation laws take the same
form~\eqref{V-SUSY} as before, the only difference being that the
fields can now depend on the position on $M$.  In components, the
formula reads
\begin{equation}
  \begin{aligned}
    \delta A_\mu &= i\lambda_\mu,
    \\
    \delta\sigma_\mu &= \lambda_\mu + V^\nu\zeta_{\nu\mu},
    \\
    \delta\lambda_\mu
    &= -iV^\nu F_{\nu\mu} + D_\mu(V^\nu\sigma_\nu),
    \\
    \delta\zeta_{\mu\nu}
    &= i F_{\mu\nu} + D_\mu\sigma_\nu - D_\nu\sigma_\mu
         - [\sigma_\mu, \sigma_\nu],
    \\
    \delta\alpha_{\mu\nu}
    &= D_\mu\sigma_\nu - D_\nu\sigma_\mu + \auxD_{\mu\nu},
    \\
    \delta \auxD_{\mu\nu}
    &= D_\mu(V^\rho\alpha_{\rho\nu}) - D_\nu(V^\rho\alpha_{\rho\mu})
         - [V^\rho\sigma_\rho, \alpha_{\mu\nu}]
         \\ &\qquad
         - D_\mu\lambda_\nu + D_\nu\lambda_\mu
         - [\lambda_\mu, \sigma_\nu]
         - [\sigma_\mu, \lambda_\nu]
         - D_\mu(V^\rho\zeta_{\rho\nu}) + D_\nu(V^\rho\zeta_{\rho\mu}).
  \end{aligned}
\end{equation}
The supersymmetry transformation laws for the chiral multiplets are
lifted to
\begin{equation}
  \begin{aligned}
    \delta\CA_m &= V^\mu\rho_{\mu m}, \\
    \delta\CAb_m &= \etab_m, \\
    \delta\rho_{\mu m} &= F_{\mu m} + iD_\mu X_m
                           + i\cD_m \sigma_\mu + V^\nu\auxF_{\nu\mu m}, \\
    \delta\etab_m
    &= V^\mu (F_{\mu m} - iD_\mu X_m)
                            + i V^\mu \cDb_m\sigma_\mu, \\
    \delta\mub_{\mu\nu m} &= \auxFb_{\mu\nu m}, \\
    \delta\auxF_{\mu\nu m}
    &= D_\mu\rho_{\nu m} - D_\nu\rho_{\mu m}
          - [\sigma_\mu, \rho_{\nu m}] + [\sigma_\nu, \rho_{\mu m}]
          - i\cD_m \zeta_{\mu\nu}, \\ 
    \delta\auxFb_{\mu\nu m}
    &= D_\mu(V^\rho\mub_{\rho\nu m}) - D_\nu(V^\rho\mub_{\rho\mu m})
          + \mub_{\mu\nu m} V^\rho \sigma_{\rho}.
  \end{aligned}
\end{equation}
% \begin{equation}
%   \begin{aligned}
%     \delta\CA &= \iota_V\rho, \\
%     \delta\CAb &= \etab, \\
%     \delta\rho &= (F_{\mu m} + iD_\mu X_m) \rmd x^\mu \wedge \rmd x^m
%                            + i\rmd_\CA \sigma + \iota_V\auxF, \\
%     \delta\etab &= \iota_V (F_{\mu m} - iD_\mu X_m) \rmd x^\mu \wedge \rmd x^m
%                             + i\rmd_\CAb \iota_V\sigma, \\
%     \delta\mub &= \auxFb, \\
%     \delta\auxF &= \rmd_A\rho - \sigma \wedge \rho - i\rmd_\CA \zeta, \\ 
%     \delta\auxFb &= \rmd_A\iota_V\mub + \mub\iota_V\sigma.
%   \end{aligned}
% \end{equation}
Typical observables are gauge-invariant operators constructed from
$\CA$, such as Wilson lines, inserted at zeros of $V$ on $\Sigma$.

Likewise, we can lift the supersymmetric action from two dimensions.
The result is
\begin{multline}
  \label{5d-S_V}
  S_V
  = \int_{ \Sigma \times M} \sqrt{h} \, \rmd^5 x \Tr\Bigl(
      \frac12 F_{\mu\nu} F^{\mu\nu}
      + D_\mu\sigma_\nu D^\mu\sigma^\nu
      +\frac{\kappa}{2} \sigma^\mu \sigma_\mu
      \\
      - \frac12 [\sigma_\mu, \sigma_\nu] [\sigma^\mu, \sigma^\nu]
      + \frac12 \auxD'_{\mu\nu} \auxD'^{\mu\nu}
      + \dotsb\Bigr)
\end{multline}
for the vector multiplet, and
\begin{multline}
  \label{5d-S_C}
  S_C = \int_{ \Sigma \times M} \sqrt{h} \, \rmd^5 x \Tr\Bigl( (F_{\mu
    m} + iD_\mu X_m + V^\nu \auxF_{\nu\mu m}) (F^{\mu m} - iD^\mu X^m
  + V^\nu \auxFb_\nu{}^{\mu m})
  \\
  + \cD_m\sigma_\mu \cDb^m\sigma^\mu
  - iD^m X_m \auxD'^{\mu\nu} \eps_{\mu\nu} + \frac12 \auxF_{\mu\nu m}
  \auxFb^{\mu\nu m}
  \\
  - i V^\nu \auxF_{\nu\mu m} \cDb^m\sigma^\mu - i V^\nu\auxFb_{\nu\mu
    m} \cD^m \sigma^\mu + \dotsb\Bigr)
\end{multline}
for the chiral multiplet.  Here $\eps_{\mu\nu}$ are components of the
volume form $\sqrt{h_\Sigma} \, \rmd^2 x$ on $\Sigma$, and we have
abbreviated boundary terms and fermionic terms.  For the
superpotential term, the form of $S_W$ is the same as in the 2d case,
with $W$ now being a gauge-invariant holomorphic functional of the
complex gauge field $\CA$ on $M$.

The action for the $\Omega$-deformed twisted 5d MSYM theory is the sum
\begin{equation}
  S = \frac{1}{2e^2} (S_V + S_C + S_W),
\end{equation}
where $e^2$ is the coupling constant of the theory.  The question is
what superpotential is the right one to use.

Note that neither $S_V$ nor $S_C$ described above contains kinetic
terms for $A_m$ and $X_m$ along $M$.  So these terms should be
generated by the superpotential.  Since the potential associated with
$\auxF$ is proportional to $|\del W/\del\CA|^2$ and the kinetic terms
for $A_m$ and $X_m$ are of second order in derivatives, $W$ must be of
first order.  A natural candidate is then the Chern--Simons functional
for $\CA$.  It turns out that the right choice is \cite{Yagi:2013fda}
\begin{equation}
  \label{CS}
  W = \frac12 \int_M \Tr\Bigl(\CA \wedge \rmd\CA 
         - \frac{2i}{3} \CA \wedge \CA \wedge \CA\Bigr).
\end{equation}
For this choice of $W$, the superpotential term \eqref{S_W} is given by
\begin{multline}
  S_W
  = \frac{i}{2} \int_{\Sigma \times M} \sqrt{h} \, \rmd^5x \Tr\Bigl(
     \frac12 \auxF_{\mu\nu l} \CF_{mn}
     + \rho_{\mu l} \cD_m \rho_{\nu n}
     + \frac12 \auxFb_{\mu\nu l} \CFb_{mn}
     + \etab_l \cDb_m \mub_{\mu\nu n}\Bigr) \eps^{\mu\nu lmn} \\
     - \frac{i}{\veps}
        \int_{\del\Sigma} W \, \rmd\vphi.
\end{multline}

To see that the above choice of $W$ is indeed the right one, set $V =
0$ and integrate out the auxiliary fields.  Integrating out $\auxD'$
gives the potential $\Tr(D_m X^m)^2$, while integrating out $\auxF$
produces $\frac12\Tr\CF_{mn}\CFb^{mn}$, where $\CF_{mn}$ are
components of the curvature of $\CA$.  Up to a total derivative on $M$
(which, for $M = \R^3$, vanishes upon integration under usual boundary
conditions), the two contributions combine to give
\begin{equation}
  \Tr\Bigl(\frac12  F_{mn} F^{mn} + D_m X_n D^m X^n
   - \frac12 [X_m, X_n] [X^m, X^n]\Bigr).
\end{equation}
After this is done, the bosonic part of the action can be written as
\begin{equation}
  \frac{1}{2e^2} \int_{\R^5} \rmd^5 x \Tr\Bigl(
  \frac12 F_{MN} F^{MN}
  + D_M X_N D^M X^N
  - \frac12 [X_M, X_N] [X^M, X^N]\Bigr)
\end{equation}
for flat spacetime $\R^5$, with $(X_M) = (\sigma_\mu, X_m)$.  This is
precisely the bosonic part of the standard 5d MSYM action.%
\footnote{The bosonic part of the undeformed action is invariant under
  phase rotations of $W$.  To fix the phase, we need to look at the fermionic
  part.  Alternatively, one can fix it by comparing the Chern--Simons
  level in our localization formula \eqref{ac-CS} with the
  identification obtained in \cite{Beem:2012mb} from a 6d point of
  view.}

Although the Chern--Simons superpotential \eqref{CS} correctly
reproduces the 5d MSYM action, it also causes a problem in the case
that $\Sigma$ has a boundary.  The real part of the Chern--Simons
functional for $\CA$ shifts by integer multiples of $2\pi$ under gauge
transformations that are not connected to the identity.  Since the
Chern--Simons functional enters the boundary term in $S_W$, this would
mean that the action is gauge invariant modulo $2\pi i$ if and only if
$1/\veps$ obeys a certain quantization condition, and otherwise the
path integral would not be well-defined.  However, we do not want to
restrict the possible values of $\veps$.  So we instead restrict the
gauge symmetry -- on the boundary, we only allow topologically trivial
gauge transformations.

The lifted formulas for $S_V$ and $S_C$ are $Q$-exact.  Hence, the
quasi-topological property of the $\Omega$-deformed theory discussed
in the 2d context still holds for the theory constructed here.  In
addition, the theory is topological on $M$, since the metric on $M$
enters the action only through $S_V$ and $S_C$.

The reader might worry that our twisted 5d MSYM theory may not be
well-defined.  Indeed, 5d gauge theories are in general not
perturbatively renormalizable by the standard argument.  Despite its
highly supersymmetric nature, 5d MSYM theory also suffers from
ultraviolet divergences starting at the six-loop level
\cite{Bern:2012di} (though there are arguments suggesting that the
theory might be rendered finite by some nonperturbative
mechanism~\cite{Douglas:2010iu, Lambert:2010iw, Lambert:2012qy}).
However, the twisted theory is an exception as one restricts attention
to the $Q$-invariant sector: one can make use of the metric
independence of the theory to shrink $\Sigma$ or $M$ to a point,
thereby reducing the theory to a lower-dimensional one which is
renormalizable.  Since this process involves a $Q$-exact deformation
of the action, it may be thought of as introduction of $Q$-exact
regulator terms.  In fact, the localization of the path integral we
are about to perform is one instance of such a reduction to a
lower-dimensional theory by a $Q$-exact deformation.  In this case,
the twisted theory is reduced to analytically continued Chern--Simons
theory.

\subsection{Localization to analytically continued Chern--Simons
  theory}
\label{5d-localiztion}

We now establish the equivalence between the $\Omega$-deformed twisted
5d MSYM theory for $\Sigma = D$ and analytically continued
Chern--Simons theory.  To this end, we view the 5d theory on $\Sigma
\times M$ as a B-twisted gauge theory on $\Sigma$, regarding $M$ as an
internal space whose coordinates are continuous ``flavor indices,''
and interpreting the integration over $M$ in the formula
\eqref{5d-S_V} etc.\ as summation over these indices.  Then we can
localize the path integral for correlation functions just as we did in
section \ref{localization}.

Recall from our discussion in section \ref{BC} that the boundary
condition for our theory is specified by a brane, whose support
$\gamma$ is a Lagrangian submanifold of the target space of the chiral
multiplet scalar.  In the present context, the scalar is the complex
gauge field $\CA$, and the target space is the space of complex
connections on $M$.  (If $M$ has a boundary, then the target space is
the space of complex connections obeying a chosen boundary condition.)
There is a natural K\"ahler form on this space, given by
\begin{equation}
  \label{omega}
  \omega = \int_M \Tr\delta A \wedge \star\delta X.
\end{equation}
The associated K\"ahler metric
\begin{equation}
  \rmd s^2 = \int_M \Tr\delta\CA \wedge \star\delta\CAb
\end{equation}
is the metric used in the construction of our chiral multiplet action
\eqref{5d-S_C}.  So we have equipped the target space with this
K\"ahler form, and $\gamma$ is a Lagrangian submanifold with respect
to it.  On the boundary, $\CA$ is required to be valued in $\gamma$.

We also need to choose the subgroup $H$ which specifies the allowed
boundary values of gauge transformations.  Previously we chose it to
be trivial, that is, we demanded gauge transformations to be trivial
on the boundary.  This time, we allow all possible gauge
transformations that preserve the boundary condition $A_n = 0$.  For
the reason explained already, they must be moreover topologically
trivial along $M$ on the boundary.  Thus, $H$ in the present case is
the group of topologically trivial $G$-gauge transformations on $M$,
which we denote by $\CG$; the corresponding moment map is
\begin{equation}
  \mu = D^m X_m.
\end{equation}
Accordingly, $\gamma$ is required to be invariant under the action of
$\CG$.

In fact, for the purpose of connecting our 5d theory to analytically
continued Chern--Simons theory, we need a stronger condition on
$\gamma$: we require
\begin{equation}
  \gamma =  \Gamma \cap \mu^{-1}(0)
\end{equation}
for some $\CG_\C$-invariant submanifold $\Gamma$ of the space of
complex connections on $M$, where $\CG_\C$ is the complexification of
$\CG$.  (As it will become clear, $\Gamma$ is identified with an
integration contour for the path integral in the Chern--Simons theory;
since this theory has invariance under complex gauge transformations,
$\Gamma$ should be invariant under $\CG_\C$, not just $\CG$.)  Due to
the K\"ahler form \eqref{omega} being only invariant under $\CG$ and
not $\CG_\C$, generically $\Gamma$ itself cannot be a Lagrangian
submanifold.  However, its restriction $\gamma$ to the $\CG$-invariant
submanifold $\mu^{-1}(0)$ can be so, and used as the support of our
brane.  The above form of $\gamma$ is compatible with our localization
condition, which actually enforces the restriction $\mu=0$.

The localization procedure is essentially the same as before.  We
deform the action as
\begin{equation}
  S \to
  t^3 S_V + tS_C + S_W
  - \frac12 \delta\int_{D \times M} \sqrt{h} \, \rmd^5x
     \Tr(t^3 \alpha^{\mu\nu}\auxD'_{\mu\nu}
  + ts\auxF^{\mu\nu m} \mub_{\mu\nu m}),
\end{equation}
rescale $\mub$ as $\mub \to s^{-1}\mub$, and send $s \to \infty$. Then
we integrate out the auxiliary fields, whereby we get $\auxF = 0$ and
the constraint \eqref{mu=Dsigma} on $\mu$.  After that, we equip $D$
with a metric with positive curvature and take $t$ to be very large to
find that the path integral localizes to the locus given by the
equations
\begin{equation}
  F_{\mu\nu} = \sigma_\mu  = F_{\mu m} = D_\mu X_m = 0.
\end{equation}
With $A_\mu$ totally gauged away, the equations become
\begin{equation}
  A_\mu = \sigma_\mu = \del_\mu A_m = \del_\mu X_m = 0.  
\end{equation}
These equations say that the nontrivial part of a localization
configuration is specified by the complex gauge field $\CA$ which must
be constant on $D$.  The brane boundary condition requires $\CA \in
\Gamma \cap \mu^{-1}(0)$, and the constraint coming from $\auxD'$
demands $\mu = 0$, which is compatible with the boundary condition.
The localization locus is therefore $\Gamma \cap \mu^{-1}(0)$.

Having identified the localization configurations, we integrate over
fluctuations around these configurations and over the fermions.  For
the gauge-fixing condition, we again use the standard one $\hh^{MN}
\nabla_M A_N = 0$, where $\hh$ can be any metric on $D \times M$.  For
us, it is convenient to use $\hh = h_D \oplus t^{3/2} h_M$, for which
the corresponding gauge-fixing term is
\begin{equation}
  \label{S_G-5d}
  S_G = t^3 \int_{D \times M} \sqrt{h} \, \rmd^5x \Tr\bigl(
           \cb \nabla^\mu D_\mu c +  t^{-3/2} \cb \nabla^m D_m c
           + (\nabla^\mu A_\mu + t^{-3/2}\nabla^m A_m)^2 \bigr).
\end{equation}
After adding this term to the action, we rescale the fluctuations and
the fermions appropriately.  The way we do this is slightly different
from the rescaling \eqref{rescaling} considered before, since this
time the ghosts have zero modes; the zero-mode parts $c_0$, $\cb_0$ of
$c$, $\cb$ are simply constants on $D$, and may be regarded as
adjoint-valued scalar fields on $M$.  (Recall that we are allowing all
gauge transformations that are compatible with the boundary condition
$A_n = 0$.)  Writing $c = c_0 + \ct$, \ $\cb = \cb_0 + \tilde\cb$, we
rescale the ghosts as
\begin{equation}
  (c_0, \ct, \cb_0, \tilde\cb)
  \to (t^{-3/4} c_0, t^{-3/2} \ct, t^{-3/4} \cb_0, t^{-3/2} \tilde\cb).
\end{equation}
The remaining fields are rescaled as before.  Noting that $c$
satisfies the boundary condition $\del_n c = 0$ just as gauge
transformation parameters do, we then find that to leading order, the
fluctuations and the fermion nonzero modes enter the action only
through terms that do not depend on the background.  Hence,
integration over them just produces a constant independent of the
background.

The final expression of the localized path integral is similar to the
formula~\eqref{loc-formula}.  Unlike the previous case, however, it
involves integration over the zero modes $c_0$, $\cb_0$.  Another
difference is that $S_G$ contains terms that depend on the background
and $c_0$, $\cb_0$:
\begin{equation}
  S_{G0} = \int_D \sqrt{h_D} \, \rmd^2 x \int_M \sqrt{h_M} \, \rmd^3x \bigl(
               \cb_0 \nabla^m D_m c_0 + (\nabla^m A_{m0})^2\bigr).
\end{equation}
Taking these differences into account, we obtain the localization
formula
\begin{equation}
  \vev{\CO}
  = \int_{\Gamma \cap \mu^{-1}(0)} \cD\CA_0 \, \cD c_0 \, \cD\cb_0 \,
     \exp(S_0 - S_{G0}) \CO,
\end{equation}
with
\begin{equation}
  S_0 = \frac{\pi i}{2e^2\veps} \int_M \Tr\Bigl(\CA_0 \wedge \rmd\CA_0 
           - \frac{2i}{3} \CA_0 \wedge \CA_0 \wedge \CA_0\Bigr).
\end{equation}

The piece $S_{G0}$ in the action that appears in the above formula may
be interpreted as a gauge-fixing term for the 3d gauge symmetry.  So
we can drop this piece if we perform the path integral over $(\Gamma
\cap \mu^{-1}(0))/\CG$.

On the other hand, the restriction of the path integral to the
subspace $\mu^{-1}(0)$ amounts to gauge fixing of the noncompact part
of the complexified gauge symmetry.  The reason is that the equation
$\mu = D^m X_m = 0$ is invariant under $\CG$ but not under $\CG_\C$,
and using $\CG_\C$, a generic complex connection can be transformed to
one that fixes $\mu = 0$.  This is actually a familiar fact about the
moduli space of vacua in supersymmetric gauge theory: the moduli space
is the zero locus of the D- and F-term potentials modulo gauge
transformations, but the same space can also be obtained by dropping
the D-term equation and taking the quotient with respect to the action
of the complexified gauge group.  Thus, we can drop the constraint
$\mu = 0$ and complexify the gauge symmetry, replacing the integration
contour with the submanfiold $\Gamma/\CG_\C$ of the moduli space $\CM$
of complex connections on $M$.  This mechanism of complexification of
the gauge symmetry has been observed previously for 5d MSYM theory on
$S^2 \times M$ \cite{Yagi:2013fda, Cordova:2013cea}.

The above formula can then be rewritten as
\begin{equation}
  \label{ac-CS}
  \vev{\CO} = \int_{\Gamma/\CG_\C} \cD\CA \exp(ik S_{\text{CS}}) \CO,
\end{equation}
where
\begin{equation}
  \label{level}
  S_{\text{CS}} = \frac{1}{4\pi} \int_M \Tr\Bigl(\CA \wedge \rmd\CA 
           - \frac{2i}{3} \CA \wedge \CA \wedge \CA\Bigr), \quad
  k = \frac{2\pi^2}{e^2\veps}.
\end{equation}
This is the path integral for Chern--Simons theory at level $k$, with
the gauge field analytically continued to a complex connection.
Therefore, we have reduced the path integral for the $\Omega$-deformed
twisted 5d MSYM theory on $D \times M$ to that for analytically
continued Chern--Simons theory on $M$, establishing the equivalence
between the two theories.

In the Chern--Simons theory, one must specify a convergent
middle-dimensional integration cycle in $\CM$.  In our localization
formula, the integration contour $\Gamma/\CG_\C$ is a Lagrangian
submanifold of $\CM$.%
\footnote{The K\"ahler form on $\CM$ is inherited from the space of
  complex connections: under the identification $\CM \simeq
  \mu^{-1}(0)/\CG$, it is represented by the restriction of the
  $\CG$-invariant two-form \eqref{omega} to $\mu^{-1}(0)$.  It
  vanishes on $\Gamma/\CG_\C \simeq (\Gamma \cap \mu^{-1}(0))/\CG$
  since $\gamma = \Gamma \cap \mu^{-1}(0)$ is a Lagrangian submanifold
  by assumption.  Being an integration cycle of the Chern--Simons
  theory, $\Gamma/\CG_\C$ is moreover middle-dimensional in $\CM$.}
A basic example of such a contour is the real contour, represented by
the space of complex connections that are $\CG_\C$-equivalent to real
connections, which is a good contour when the Chern--Simons level is
real.

\section{3d-3d correspondence}
\label{3d-3d}

To conclude our discussion, in the final section we interpret the
results we obtained about the $\Omega$-deformed twisted 5d MSYM theory
from the point of view of the 3d-3d correspondence.  This allows us to
establish the correspondence between the 3d $\CN = 2$ superconformal
theory $T[M]$ and analytically continued Chern--Simons theory on $M$.
Furthermore, we will see that our construction of the 5d theory,
together with the 3d-3d correspondence, implies a mirror symmetry
between $\Omega$-deformed 2d theories.

\subsection[\texorpdfstring{$T[M]$}{T[M]} and analytically continued
Chern--Simons theory]{$\boldsymbol{T[M]}$ and analytically continued
Chern--Simons theory}

Consider the $(2,0)$ theory on $S^1 \times_V \Sigma \times M$, with
$S^1$ a circle of radius $R$ and $V$ a Killing vector field on
$\Sigma$.  Here, the space $S^1 \times _V \Sigma$ is a nontrivial
$\Sigma$-fibration over $S^1$, constructed from the trivial fibration
$[0, 2\pi R] \times \Sigma$, by gluing the two ends of the interval
$[0, 2\pi R]$ with an action of the isometry $\exp(2\pi RV)$ on the
fiber $\Sigma$.  The structure group of the spinor bundle of this
space is reduced to $\Spin(2)_\Sigma \times \Spin(3)_M$, and the
R-symmetry group of the theory is $\Spin(5)_R$.  This is just like the
case of 5d MSYM theory on $\Sigma \times M$.  Thus, we can consider
topological twisting analogous to the one applied to that theory.

It is well known that for flat spacetime, the $(2,0)$ theory
compactified on $S^1$ is equivalent, at low energies, to 5d MSYM
theory with gauge coupling $e^2 = 4\pi^2 R$.  In view of this
relation, we propose that at energies much smaller than $1/R$, the
above twisted $(2,0)$ theory on $S^1 \times_V \Sigma \times M$ is
equivalent to the $\Omega$-deformed twisted 5d MSYM theory on $\Sigma
\times M$ constructed in the previous section, with the same gauge
coupling and the $\Omega$-deformation given by a Killing vector field
proportional to $V$.

Another regime that is relevant to us is the one in which energies are
much smaller than $1/L$, where $L$ is the length scale of $M$.  In
this regime, the $(2,0)$ theory compactified on $M$ gives $T[M]$ by
definition.  Hence, the twisted $(2,0)$ theory reduces to a
topologically twisted version of $T[M]$ on $S^1 \times_V \Sigma$.

Based on our proposal and this observation, we can show that the
$\Omega$-deformed twisted 5d MSYM theory is equivalent to the twisted
$T[M]$.  The argument goes as follows.

We fix an energy scale $E$, and consider the twisted $(2,0)$ theory on
$S^1 \times_V \Sigma \times M$ with $R$, $L \ll 1/E$.  This theory can
be described either as the $\Omega$-deformed twisted 5d MSYM theory on
$\Sigma \times M$, with $e^2$ and $M$ small, or as the twisted $T[M]$
on $S^1 \times_V \Sigma$, with the $S^1$ small.  The 5d theory is
topological on $M$, so we can scale up $M$ if we wish.  Likewise, the
3d theory is independent of $R$ and we can set it to any value as long
as we keep unchanged the isometry $\exp(2\pi RV)$ (and other possible
fugacity parameters associated to boundaries in $M$), for correlation
functions on $S^1 \times_V \Sigma$ are supersymmetric indices.  (See
e.g.~\cite{Beem:2012mb} for more discussions on this point.)

The last statement suggets that the 5d theory depends on $e^2$ only
through the combination $e^2 V$, and this is indeed true.  To see
this, we consider a $Q$-exact deformation of the action similar to the
one used in the derivation of the localization formula for $\Sigma =
D$ in section~\ref{5d-localiztion}.  After such a $Q$-exact
deformation, only $S_V$, $S_C$ and the boundary term in $S_W$ are
relevant for the computation of the path integral.  The claim then
follows from the fact that the dependence on $e^2$ coming from the
first two is $Q$-exact, while the boundary term of the action depends
on $e^2$ through the factor $1/e^2\veps$.  Thus, we can rescale $e^2$
to any value, if we simultaneously rescale $V$ to keep $e^2V$ fixed.

\begin{figure}[h]
  \centering
  \begin{tikzpicture}
    \node[draw] (6d) at (0,0) {$(2,0)$ theory on $S^1 \times_V \Sigma \times M$};
    \node[draw] (5d) at (-3,-2) {$\Omega$-def'd 5d MSYM on $\Sigma \times M$};
    \node[draw] (3d) at (3, -2) {$T[M]$ on $S^1 \times_V \Sigma$};

    \draw[->] (6d) to (5d);
    \draw[->] (6d) to (3d);
    \draw[<->] (5d) to (3d);
  \end{tikzpicture}

  \caption{Equivalence between the $\Omega$-deformed twisted 5d MSYM
    theory and the twisted~$T[M]$}
  \label{6d-5d-3d}
\end{figure}
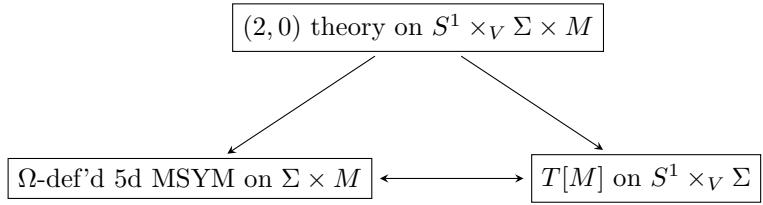

Since the 5d and 3d theories are different descriptions of the same 6d
theory, they are equivalent, and this is valid at any energy scale
$E$, for any values of $e^2$ and $R$, and for any metric on $M$.
Therefore, we conclude that the $\Omega$-deformed twisted 5d MSYM
theory on $\Sigma \times M$ is equivalent to the twisted $T[M]$ on
$S^1 \times_V \Sigma$.  Our argument is depicted in
figure~\ref{6d-5d-3d}.

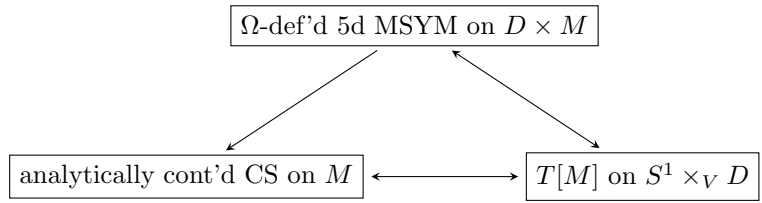
\begin{figure}[h]
  \centering
  \begin{tikzpicture}
    \node[draw] (5d) at (0,0) {$\Omega$-def'd 5d MSYM on $D \times M$};
    \node[draw] (CS) at (-3,-2) {analytically cont'd CS on $M$};
   \node[draw] (3d) at (3,-2) {$T[M]$ on $S^1 \times_V D$};

    \draw[->] (5d) to (CS);
    \draw[<->] (5d) to (3d);
    \draw[<->] (CS) to (3d);
  \end{tikzpicture}

  \caption{Correspondence between $T[M]$ and analytically continued
    Chern--Simons theory}
  \label{5d-CS-3d}
\end{figure}

Now we take $\Sigma = D$.  In this case we have shown that the
$\Omega$-deformed twisted 5d MSYM theory is equivalent to analytically
continued Chern--Simons theory.  Combined with the equivalence just
discussed, this establishes the correspondence between $T[M]$ and the
latter theory (figure~\ref{5d-CS-3d}).

Let us briefly comment on an alternative explanation for this
correspondence, proposed by Beem et al.~\cite{Beem:2012mb}.  Their
approach starts with the same 6d setup as ours, namely the $(2,0)$
theory on $S^1 \times_V D \times M$.  The main difference is that in
their case, in addition to reduction on the $S^1$, one considers
deforming $D$ to a cigar shape and reducing the theory on the circle
fibers of $D$.  After doing so, one has a twisted $\CN = 4$ super
Yang--Mills theory on the product of an interval and $M$.  Then one
can invoke an argument given in~\cite{Witten:2010zr, Witten:2011zz}
and show that the system is equivalent to the Chern--Simons theory.
Our derivation has the advantage that it avoids questions concerning
the singular point of the geometry, that is the tip of the cigar,
where the circle fiber shrinks to a point and the analysis becomes
difficult.

In deriving the correspondence between $T[M]$ and analytically
continued Chern--Simons theory, we set $\Sigma = D$ and impose
boundary conditions of a specific type.  Similar localization
computations may be carried out for other choices of $\Sigma$ and
boundary conditions, and may lead to yet unknown correspondences.

\subsection[\texorpdfstring{$\Omega$}{Omega}-deformed mirror
symmetry]{$\boldsymbol\Omega$-deformed mirror symmetry}

The equivalence between the $\Omega$-deformed twisted 5d MSYM theory
and the twisted $T[M]$ implies more than just the correspondence
discussed above.  We can use it to find another interesting
correspondence which relates two $\Omega$-deformed 2d theories.

Consider 5d MSYM theory, compactified and topologically twisted on
$M$.  In the limit where $M$ is very small, it becomes an $\CN =
(2,2)$ theory $\Tt[M]$ in two dimensions.  An analysis along the lines
of \cite{Bershadsky:1995vm} shows that $\Tt[M]$ is a Landau--Ginzburg
model whose target space is the moduli space $\CM_{\text{flat}}$ of
complex flat connections on $M$, assuming that the flat connections
are irreducible.%
\footnote{In general, the Landau--Ginzburg model description breaks
  down at reducible flat connections due to appearance of extra
  massless modes on $M$ coming from $A_\mu$, $\sigma_\mu$ and their
  superpartners.  This echoes the observation made in
  \cite{Chung:2014qpa, Dimofte:2014zga} that the construction of
  $T[M]$ proposed in \cite{Dimofte:2011ju, Dimofte:2013iv} really
  captures only the subsector of the full theory, obtained by
  truncation to the irreducible connections.}

If we instead start from the $\Omega$-deformed twisted 5d MSYM theory
on $\Sigma \times M$, then we obtain an $\Omega$-deformed, twisted
version of $\Tt[M]$ on $\Sigma$.  The model is more precisely
B-twisted, as our construction of the 5d theory is based on a
B-twisted gauge theory, and the chiral multiplets of the model simply
come from their counterparts in the 5d theory, containing $\CA_m$.
Alternatively, one may note that generically $\U(1)_V$ would be broken
by the superpotential, so the twisting should be done with $\U(1)_A$.
(If the model happens to have a quasi-homogeneous superpotential, one
can deform the 5d theory so that nonhomogenous terms are generated;
then one knows that the 2d theory is B-twisted, as the twisting does
not change under such a deformation.)

On the other hand, $T[M]$ compactified on $S^1$ reduces to an $\CN =
(2,2)$ theory $\Th[M]$ in the limit $R \to 0$.  So if we instead start
with the twisted version of $T[M]$ formulated on $S^1 \times_V
\Sigma$, then we get an $\Omega$-deformed twisted $\Th[M]$ on
$\Sigma$.

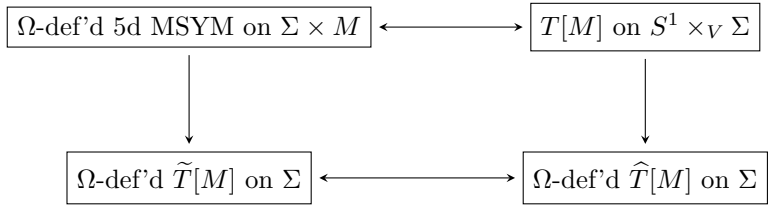
\begin{figure}[h]
  \centering
  \begin{tikzpicture}
    \node[draw] (5d) at (-3,0) {$\Omega$-def'd 5d MSYM on $\Sigma \times M$};
    \node[draw] (3d)  at (3,0) {$T[M]$ on $S^1 \times_V \Sigma$};
    \node[draw] (2d-B)  at (-3,-2) {$\Omega$-def'd $\Tt[M]$ on $\Sigma$};
    \node[draw] (2d-A)  at (3,-2) {$\Omega$-def'd $\Th[M]$ on $\Sigma$};

    \draw[<->] (5d) to (3d);
    \draw[->] (5d) to (2d-B);
    \draw[->] (3d) to (2d-A);
    \draw[<->] (2d-A) to (2d-B);
  \end{tikzpicture}

  \caption{$\Omega$-deformed mirror symmetry}
  \label{5d-2d-3d}
\end{figure}

Now, combining the facts that (1) the $\Omega$-deformed twisted 5d
MYSM theory is topological on $M$; (2) the twisted $T[M]$ on $S^1
\times_V \Sigma$ is independent of $R$ (as long as $RV$ and other
fugacities are fixed); and (3) these two theories are equivalent, we
deduce that the $\Omega$-deformed twisted $\Tt[M]$ is equivalent to
the $\Omega$-deformed twisted $\Th[M]$ (figure~\ref{5d-2d-3d}).

This equivalence may be thought of as a mirror symmetry.  The reason
is that while the twisted 5d MSYM theory reduced on $M$ gives rise to
a B-twisted Landau--Ginzburg model, reduction of the twisted $T[M]$ on
the $S^1$ produces an \emph{A-twisted} gauge theory, if $T[M]$ is
realized as gauge theory as in \cite{Dimofte:2010tz, Dimofte:2011ju};
in particular, it can flow to an A-twisted sigma model in the
infrared.  This may be seen from the fact that a scalar in the vector
multiplet of the 2d theory comes from a component of the 3d gauge
field, which is neutral under the R-symmetry $\U(1)_R$ used in the
topological twist of the 3d theory.  Since the scalar is charged under
the axial R-symmetry $U(1)_A$, it follows that $\U(1)_R$ becomes the
vector R-symmetry $\U(1)_V$.

\begin{figure}[h]
  \centering
  \begin{tikzpicture}
    \node[draw] (CS) at (0,0) {analytically cont'd\ CS on $M$};
    \node[draw] (2d)  at (-3,-2) {$\Omega$-def'd B-tw'd $\Tt[M]$ on $D$};
    \node[draw] (3d)  at (3,-2) {$\Omega$-def'd A-tw'd $\Th[M]$ on $D$};

    \draw[<->] (CS) to (2d);
    \draw[<->] (CS) to (3d);
    \draw[<->] (2d) to (3d);
  \end{tikzpicture}

  \caption{A triangle of correspondences}
  \label{CS-2d-3d}
\end{figure}
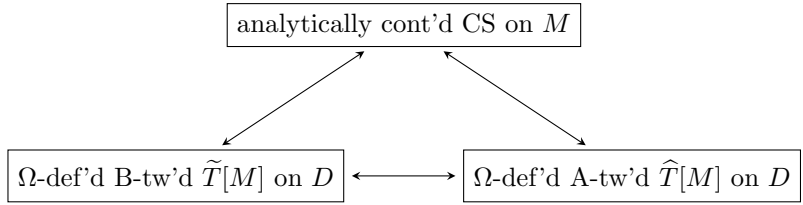

Specializing to the case $\Sigma = D$, we can place the correspondence
between $T[M]$ and analytically continued Chern--Simons theory
(figure~\ref{5d-CS-3d}) and the one between $\Tt[M]$ and $\Th[M]$
(figure~\ref{5d-2d-3d}) in a single diagram (figure~\ref{CS-2d-3d}).  The
result is an intriguing triangle of correspondences that connects
analytically continued Chern--Simons theory, $\Tt[M]$ and $\Th[M]$.

Using the relation between $\Tt[M]$ and analytically continued
Chern--Simons theory, we can extract information on the superpotential
$\Wt$ of $\Tt[M]$ as follows.

Integration cycles for the Chern--Simons theory are described by Morse
theory, with the real part of ($i$ times) the Chern--Simons action
$ikS_{\text{CS}}$ taken as the Morse function~\cite{Witten:2010cx}.
To obtain a good integration cycle, one picks a middle-dimensional
submanifold $\CCt$ of $\CM_{\text{flat}}$, and considers downward flow
lines that start from some point on $\CCt$; let
$\CC_{\CA_{\text{flat}}}$ denote the set of such lines starting from a
flat connection $\CA_{\text{flat}}$.  (If $\CM$ has components of
different dimensions, $\CCt$ is middle-dimensional in each component
of fixed dimension.)  Then, $\CC = \bigcup_{\CA_{\text{flat}} \in
  \CCt} \CC_{\CA_{\text{flat}}}$ represents a desired integration
cycle: it is middle-dimensional in the moduli space $\CM$ of complex
connections, and the path integral is convergent over it since
$\Re(ikS_{\text{CS}})$ decreases along the flow lines.  Given an
integration contour $\CC$ constructed in this manner, one can compute
the partition function by performing the path integral first over
$\CC_{\CA_{\text{flat}}}$, and then over all possible starting points
$\CA_{\text{flat}}$.  The first step defines a function $f$ on $\CCt$,
with which the partition function can be written as
\begin{equation}
  \label{Z-CS}
  Z = \int_{\CCt} f \, \rmd\CA_{\text{flat}},
\end{equation}
where $\rmd\CA_{\text{flat}}$ is a holomorphic volume form on
$\CM_{\text{flat}}$.

For the Chern--Simons theory obtained in our setup, the integration
contour $\CC$ is represented by the submanifold $\Gamma/\CG_\C$ of
$\CM$ which determines the support of the brane in the 5d theory.
This submanifold is Lagrangian, not only middle-dimensional.  When
$\CC$ has this property, $\CCt$ is represented by a Lagrangian
submanifold of $\CM_{\text{flat}}$.  Then, $\CCt$ naturally defines
the support of a brane for $\Tt[M]$ whose target space is
$\CM_{\text{flat}}$, and the partition function of $\Tt[M]$ on $D$ in
the presence of this brane is given by \cite{Yagi:2014toa}
\begin{equation}
  Z = \int_\CCt \rmd\CA_{\text{flat}}
         \exp\Bigl(\frac{2\pi i}{\veps}  \Wt\Bigr).
\end{equation}
According to the correspondence we found above, this is to be
identified with the partition function \eqref{Z-CS} of the
Chern--Simons theory.  Comparing the two expressions, we see
\begin{equation}
  f = \exp\Bigl(\frac{2\pi i}{\veps} \Wt\Bigr).
\end{equation}
Hence, information on $\Wt$ can be extracted by computing the
partition function of the Chern--Simons theory over appropriate
integration cycles.

\section*{Acknowldgments}

We would like to thank Meer Ashwinkumar for helpful discussions.  The
work of J.Y. is supported by INFN Postdoctoral Fellowship and INFN
Research Project ST\&FI.  The work of Y.L., M.-C.T. and Q.Z. is
supported by NUS Tier 1 FRC Grant R-144-000-316-112.

\bibliography{../junya}{}
\bibliographystyle{JHEP}
\end{document}